\documentclass[journal]{IEEEtran}

\usepackage{enumerate}
\usepackage{graphicx}
\usepackage{epsf}
\usepackage{subfigure}
\usepackage{amsmath}
\usepackage{amssymb}
\usepackage{array}
\usepackage{setspace}
\usepackage[amsmath,thmmarks]{ntheorem}
\usepackage[ruled,lined]{algorithm2e}
\usepackage{algorithmic}
\usepackage{color}
\usepackage{amsfonts}
\usepackage{dsfont}
\usepackage{xfrac}
\usepackage{breqn}
\usepackage{bbm}
\usepackage{cite}
\usepackage{epstopdf}
\usepackage{mathtools}

\graphicspath{{Fig/},{fig/}}

\theoremseparator{.}

\newcommand{\tabincell}[2]{\begin{tabular}{@{}#1@{}}#2\end{tabular}}

\DeclareMathOperator*{\argmax}{\arg\!\max}

\DeclareMathOperator{\diag}{diag}

\begin{document}

\title{\huge Aerial Multi-Functional RIS in Fluid Antennas-Aided Full-Duplex Networks: A Self-Optimized Hybrid Deep Reinforcement Learning Approach}

\author{
Li-Hsiang Shen,~\IEEEmembership{Member,~IEEE},
Yu-Quan Zheng
}

\maketitle

\begin{abstract}
To address the escalating data traffic demands of sixth-generation (6G) wireless networks, this paper proposes a novel architecture that integrates autonomous aerial vehicles (AAVs) and multi-functional reconfigurable intelligent surfaces (MF-RISs), termed as AM-RIS, in fluid antenna (FA)-assisted full-duplex (FD) network serving simultaneous uplink (UL) and downlink (DL) users. The AM-RIS provides hybrid functionalities, including signal reflection, amplification, and energy harvesting (EH), potentially improving both signal coverage and sustainability. Meanwhile, FA facilitates fine-grained spatial adaptability at FD-enabled base station (BS), which complements residual self-interference (SI) suppression. We aim at maximizing the overall energy efficiency (EE) by jointly optimizing transmit DL beamforming at BS, UL user power, configuration of AM-RIS (amplification, phase-shifts, and EH coefficients), and positions of the FA and AM-RIS. Owing to the hybrid continuous-discrete parameters and high dimensionality of the intractable problem, we have conceived a self-optimized multi-agent hybrid deep reinforcement learning (DRL) framework (SOHRL), which integrates multi-agent deep Q-networks (DQN) and multi-agent proximal policy optimization (PPO), respectively handling discrete and continuous actions. To enhance self-adaptability, an attention-driven state representation and meta-level hyperparameter optimization are incorporated into SOHRL, enabling multi-agents to prioritize critical features and autonomously adjust learning hyperparameters. Simulation results validate the effectiveness of the proposed AM-RIS-enabled FA-aided FD networks empowered by the conceived SOHRL algorithm. The results reveal that SOHRL outperforms existing benchmarks of the case without attention mechanism and conventional hybrid/multi-agent/standalone DRL. Moreover, the proposed AM-RIS architecture in FD achieves the highest EE performance compared to half-duplex, conventional rigid antenna arrays, partial EH, and conventional RIS without amplification, highlighting its potential as a compelling solution for EE-aware wireless networks.
\end{abstract}

\begin{IEEEkeywords}
Multi-functional RIS, autonomous aerial vehicles, full-duplex, fluid antennas, deep reinforcement learning.
\end{IEEEkeywords}

{\let\thefootnote\relax\footnotetext
{Li-Hsiang Shen and Yu-Quan Zheng are with the Department of Communication Engineering, National Central University, Taoyuan 320317, Taiwan (email: shen@ncu.edu.tw, wan900627@gmail.com). (Corresponding Author: Li-Hsiang Shen)}
}

\section{Introduction}

\subsection{Background and Literature Study}

In the era of unprecedented wireless data growth and the evolution toward the sixth-generation (6G) networks, wide-coverage, high-rate, energy-efficient, and flexible communications are becoming critical \cite{1, add1}. Reconfigurable intelligent surfaces (RIS) have been introduced as low-cost, programmable metasurfaces capable of manipulating incident signals through passive signal controls. Unlike conventional relays, RIS can perform instant passive beamforming to reconfigure wireless environments with low power consumption by steering signals toward intended users while mitigating interference \cite{2, ris1, ris2, my1, my100, my101}. However, conventional RIS is limited by its half-space coverage, passive-only capability, and depends on external power sources, which restrict its flexibility and deployment scenarios, particularly under limited energy and mobile environments. To overcome these limitations, the novel architecture of multi-functional RIS (MF-RIS) has been proposed \cite{mfris1, my7}, extending conventional reflection-only RIS by integrating active components for signal amplification and energy harvesting (EH) modules that collect ambient signal energy to enable self-sustainable operation \cite{mfris1}. This hybrid capability allows MF-RISs to enhance both coverage area and energy efficiency (EE), making it particularly suitable for applications involving mobile devices and aerial vehicles.

Moreover, the deployment of RIS and MF-RIS becomes a compelling necessity in complex and dynamic environments. In paper \cite{my_AI}, a prototype demonstration of optimal and adaptive deployment of autonomous vehicle-aided RISs is presented; however, its applicability remains constrained to short-range terrestrial mobility. Among the promising enablers, autonomous aerial vehicles (AAVs) \cite{uav2} have emerged as resilient deployment capable of carrying RISs/MF-RISs for rapid deployment, dynamic coverage and virtual line-of-sight (LoS) link establishment. By leveraging flexible spatial mobility, AAVs with RIS can function as intelligent relays, effectively handling scenarios such as emergencies, disasters, and the areas without deployment of the base station (BS) as a complementary on-demand connectivity \cite{my7, uav3, uav9}. In paper \cite{my7}, it emphasizes the benefits of incorporating RISs and AAVs for supporting LoS links. The work \cite{uav3} provides the optimal deployment strategy of aerial RIS for physical layer security. The authors in \cite{uav9} have embedded RIS on the AAV to improve both EE and coverage while fulfilling the minimum data rate requirements. Furthermore, integrating MF-RISs with AAVs offers additional sustainability in addition to the signal reflection advantages of RIS capability, i.e., MF-RIS can compensate the power utilization of mechanical flight of AAVs. In the work of \cite{my7}, the results have validated the EE performance enhancement achieved by leveraging MF-RIS in the non-terrestrial networks. The authors in \cite{uav4} analyze the optimal flying position of AAVs equipped with MF-RIS applying for covert communications. Therefore, it becomes compelling imperative to incorporate MF-RISs onto AAVs in support of large-scale wireless networks.

In addition to enhancing EE, improving spectral efficiency (SE) is also essential. Full-duplex (FD) communication significantly boosts SE by allowing simultaneous uplink (UL) and downlink (DL) transmissions over the same frequency, theoretically doubling the SE compared to the conventional half-duplex (HD) systems. However, FD operation introduces significant challenges due to self-interference (SI), particularly when strong reflected DL signals impair the UL reception at the BS, and when UL transmissions generate interference at the DL user terminals \cite{hyperparam, FD1, FD2}. The integration of AAV-mounted RISs/MF-RISs facilitates SI suppression in dynamic and highly reflective environments \cite{FD-RIS1, FD-RIS2, FD-RIS3, FD-RIS4}. Specifically, the AAV provides spatial flexibility to balance large-scale distance-dependent fading between UL and DL directions, while the RIS/MF-RIS dynamically configures amplitude and phase-shifts to align channel coefficients for mitigating SI. In \cite{FD-RIS1}, RIS on AAV is proposed for alleviating the residual SI under the use of non-orthogonal multiple access in FD transmissions. The work in \cite{FD-RIS2} considers energy-aware trajectory design for AAV-mounted RISs and relays in an FD network, aiming at enhancing user fairness constrained by the available on-board energy. However, the effectiveness of SI cancellation using AAV-mounted RIS or MF-RIS is confined by their physical separation from the FD-empowered transceivers. The SI originating directly from the transceiver is typically much stronger than the reflected signals from the RIS or MF-RIS, thereby making it challenging to fully eliminate it through passive reflection alone.

Therefore, fluid antenna (FA) systems have attracted growing attention for their powerful ability to dynamically reposition antenna elements in response to small-scale fading and interference cancellation \cite{fa1, fa2, fa3, fa4, my4, lim}. When mounted on the BS, FA array-enhanced systems introduce fine-grained spatial adaptability that complements the large-scale trajectory flexibility of AAVs, whereas the transmit beamforming of antennas complements the configurations of MF-RISs, which are not mentioned in most recent works. This structure enables enhanced multi-beamforming, reduced interference, and energy-efficient transmission under rapidly changing wireless conditions, such as vehicular networks, indoor deployments and dense urban areas. Related studies on FA systems have demonstrated the necessity and benefits of employing FA over conventional fixed antenna arrays \cite{fa5}. In particular, the work in \cite{my4} further shows that even utilizing only a subset of FA, such as half of the array, can achieve higher data rates compared to the conventional fixed arrays. As a complementary solution for FD-enabled transceivers, FA can further suppress residual SI, augmenting the SI mitigation capabilities of AAVs and MF-RISs. This is attributed to its ability to physically separate the transmitter and receiver antennas at BS over small spatial scales, thereby reducing near-field coupling and improving isolation \cite{ma1, ma2}. 
The work in \cite{ma1} considers point-to-point SI cancellation by dynamically repositioning transmit/receive antennas with minimum achievable rate guaranteed. The paper of \cite{ma2} investigates optimization of FD antenna positions in order to serve users and improve the overall secrecy rates. While existing studies have examined the deployment of FA at either mobile terminals or BSs, their joint integration with MF-RIS and AAV-assisted environments remains unexplored. However, optimizing FA placement alongside AAV paths, MF-RIS configurations, and transmit beamforming vectors adds significant complexity to the joint design problem, which should be addressed. 

Given the high dimensionality, non-convexity, and complex nature of control variables, including BS beamforming, AAV positions and configurations of MF-RIS, traditional model-based optimization approaches are insufficient. Also, these classical approaches often rely on strong assumptions about instantly attained channel models and network topology, which do not hold in highly dynamic environments. Accordingly, deep reinforcement learning (DRL) offers an intelligent system that agents can learn optimal strategies through sequential interaction with the wireless environment \cite{my_AI, rl1, rl2, rl_wu, fd_drl1}. By directly learning from environmental feedback and adapting to new scenarios, DRL can perform real-time optimization without requiring complete models of the network or channel.

\subsection{Motivation and Contributions}
In this work, we explore a novel EE-aware communication architecture that integrates AAVs, FAs, and MF-RISs, jointly forming the aerial MF-RIS (AM-RIS) system operating in FD networks. Despite its promising capabilities, the proposed AM-RIS framework faces several fundamental challenges. First, the high mobility of AAVs and their potential aerial topology result in rapidly changing channels, whereas maintaining stable links becomes difficult due to fluctuations and varying interference. Additionally, AM-RIS introduces hybrid configurations of signal reflection, amplification, and EH, potentially complicating the system with additional nonlinear constraints and coupling parameters. Moreover, FD operation introduces inevitable strong SI at BS, which is further aggravated by AM-RIS amplification and reflections, which require coordinating both transmit beamforming and AM-RIS configurations for suppressing effective SI. The deployment of FAs at BS introduces a new degree of spatial flexibility at a comparably smaller scale than AAVs \cite{my4}. However, it presents crucial challenges in terms of optimizing FD antenna positions to balance channel alignment, coverage and interference mitigation under a dynamic environment. Furthermore, the need for joint optimization over AAV trajectory, FA position, AM-RIS configurations, and transmit beamforming under limited power budget, interference, and rate requirements poses an extremely high-dimensional and non-convex decision space, which is unsolvable by using conventional optimization frameworks.

To elaborate a little further, the decision variables include both discrete domains and continuous domains, which cannot be efficiently handled by a standalone DRL algorithm. In this context, a hybrid DRL framework should be designed to handle hybrid parameters in general cases \cite{HyRL_StarRIS, my_AI, chimera}. However, conventional hybrid DRL methods potentially suffer from slow convergence, high variance, and unstable policy updates under high-dimensional state representations and partial observations in highly dynamic environment. These limitations motivate the employment of attention-based mechanism \cite{atten} that dynamically prioritizes informative features associated with critical channel conditions, thus guiding the learning process toward highly impacted regions in the state space. Furthermore, the dynamic environment necessitates self-optimizing and adaptive mechanisms capable of autonomously fine-tuning hyperparameters over time \cite{hyperparam, rl_wu}, rather than relying on manually fixed parameter settings. These enhancements intelligently enable the system to adaptively coordinate distributed agents, and significantly improve the EE performance of AM-RIS and FA-empowered FD networks. The main contributions of this work are elaborated as follows.
\begin{itemize}

\item We propose a novel architecture of AM-RIS by integrating AAVs and MF-RISs in support of FA-assisted FD networks, which enables simultaneous UL and DL transmissions. Each AM-RIS is capable of reflecting, amplifying, and harvesting energy from wireless signals in a self-sustainable manner. Note that the FA array is equipped on the FD-BS. The joint small- and large-scale position adjustments of FA and of AM-RIS significantly enhance service coverage area in dynamic environments.

\item We aim for maximizing system EE by determining DL beamforming at BS, allocated UL user power, AM-RIS configurations of amplitude/phase/EH ratios, as well as the position adjustments of FA at BS and of AM-RISs. The problem accounts for the constraints of power budget at BS and AM-RIS, minimum UL/DL user rates, as well as available deployment boundaries of the FA and AM-RISs.

\item To solve the complex problem, we conceive a self-optimized hybrid DRL (SOHRL) algorithm that integrates multi-agent deep Q-network (DQN) \cite{dqn} and multi-agent proximal policy optimization (PPO) \cite{ppo} for handling discrete and continuous parameters, respectively. To further improve the system performance, SOHRL incorporates the attention-based learning and self-optimization of hyperparameter updates, empowering agents to efficiently and effectively coordinate and interact with the dynamic environment.

\item Simulation results have validated the effectiveness of the proposed AM-RIS-enabled FA-aided FD networks with SOHRL algorithm under a variety of parameter settings. The results demonstrate that SOHRL outperforms the existing benchmarks, including standalone DQN, PPO, multi-agent DRL, and the SOHRL without attention and self-optimization mechanisms. The proposed architecture of AM-RIS for FA-aided FD also accomplishes the highest EE performance compared to the HD transmissions, fixed antenna arrays, partial EH and conventional RIS without amplification capability, highlighting its potential as a compelling solution for EE-aware wireless networks.
\end{itemize}


\section{System Model and Problem Formulation} \label{sec_sys}

\subsection{System Architecture} 

\begin{figure}[!t]
	\centering
	\includegraphics[width=3.3in]{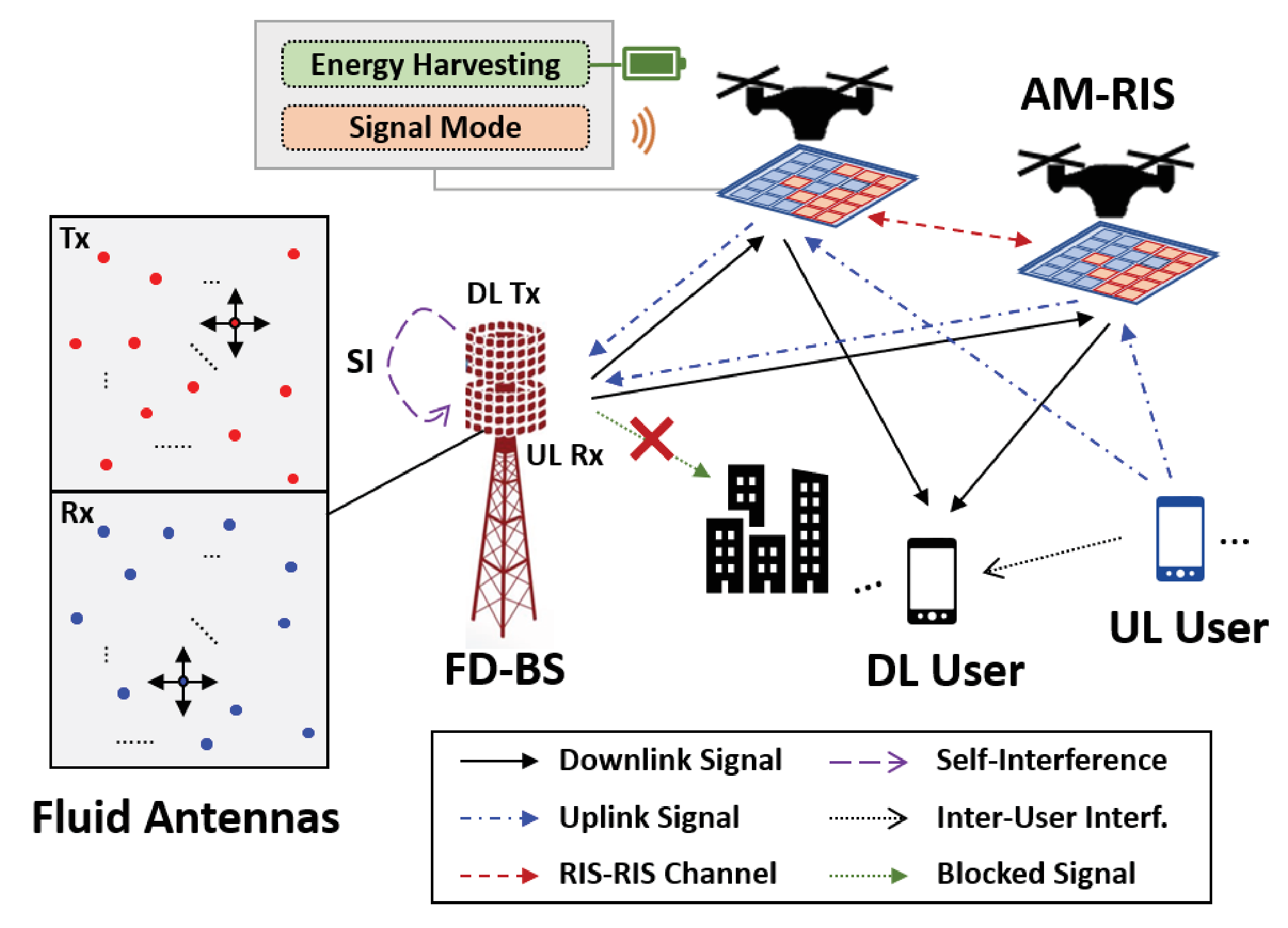}
	\caption{The proposed architecture of multi-AM-RISs for the FA-aided FD network, where FD-BS serves HD-DL/HD-UL users. The AM-RIS is implemented on AAV via MF-RIS. Note that simultaneous UL/DL transmissions occur in the cell. The DL users receive inter-DL and AM-RIS-induced reflected UL user interferences, whereas the BS Rx receives direct and AM-RIS-induced SI from DL Tx antennas. Both DL users and BS Rx receive AM-RIS amplified noise.}
	\label{systemModel}
\end{figure}

In Fig.~\ref{systemModel}, we consider an FA-empowered FD-BS having $N_{\rm T}$ transmit (Tx) and $N_{\rm{R}}$ receiving (Rx) antennas with the respective sets of $\mathcal{N}_{\rm T} = \{1,2,...,N_{\rm T}\}$ and $\mathcal{N}_{\rm R} = \{1,2,...,N_{\rm R}\}$, serving $K_{\rm D}$ downlink and $K_{\rm U}$ uplink users with respective sets of $\mathcal{K}_{\rm D} = \{1,2,...,K_{\rm D}\}$ and $\mathcal{K}_{\rm U} = \{1,2,...,K_{\rm U}\}$. Users are in the HD mode, with each equipped with a single antenna. The size of FA arrays of both Tx and Rx are $L = L_x \cdot L_y$, where $L_x$ and $L_y$ are the lengths of horizontal and vertical directions, respectively. Note that the FA receiver panel is positioned adjacent to the transmitter panel, with its center displaced by $L_x$ in the x-direction. There exist $I$ AM-RISs with the set of $\mathcal{I}=\{1,2,...,I\}$, where each AM-RIS has $M$ elements in conjunction with its set of $\mathcal{M}=\{1,2,...,M\}$. Note that $M = M_x \cdot M_y$, where $M_x$ and $M_y$ are the numbers of elements at horizontal and vertical directions, respectively. We consider a duration of $\tau$ for each timeslot, and the network configuration is supposed to remain static within each timeslot.

The three-dimensional (3D) Cartesian coordinate is considered with the locations of the BS, AM-RIS, and users defined as $\mathbf{X}^{\rm BS}= [ x^{\rm BS}, y^{\rm BS}, z^{\rm BS}]^{\mathcal{T}}$, $\mathbf{X}_{i}=[x_i, y_i, z_i]^{\mathcal{T}}$, and $\mathbf{X}^{\rm UE}_{k}=[x^{\rm UE}_k, y^{\rm UE}_k, z^{\rm UE}_k]^{\mathcal{T}}$. Note that the position of MF-RIS in AM-RIS is regarded the same as that of the AAV. Due to the practical limited coverage, the deployable region of AM-RIS $l$ is constrained by $\mathbf{X}_{\rm min} \preceq \mathbf{X}_{i}(t) \preceq \mathbf{X}_{\rm max}$, i.e., $x_{\rm min}\leq x_i \leq x_{\rm max}$, $y_{\rm min}\leq y_i \leq y_{\rm max}$, and $z_{\rm min}\leq z_i \leq z_{\rm max}$. Moreover, aerial safety of preventing AM-RIS collision is also considered, which is constrained by $d_{i,j} \triangleq \left\lVert \mathbf{X}_{i} - \mathbf{X}_{j} \right\rVert \geq d_{\rm min}$. The velocity of AM-RIS is defined as $\mathbf{v}_{i} = [v_{x,i}, v_{y,i}, v_{z,i}]^{\mathcal{T}}$, with
its initial position as $\mathbf{X}_{i}(0)$. Then we can model the flight of the AM-RIS by setting the following position and velocity constraints, given by \cite{uav_move1}
\begingroup
\allowdisplaybreaks
\begin{subequations} \label{AAV_con}
\begin{align}
	& \mathbf{X}_{i} = \mathbf{X}_{i}^{\text{o}} + \mathbf{v}_{i} \tau, \quad \forall i\in \mathcal{I}, \label{uav_con1}
	\\
	& \mathbf{X}_{\rm min} \preceq \mathbf{X}_{i} \preceq \mathbf{X}_{\rm max}, \quad \forall i\in \mathcal{I},  \label{uav_con2}
	\\
	& \left\lVert \mathbf{v}_{i} \right\rVert \leq v_{\rm max}, \quad \forall i\in \mathcal{I}, \label{uav_con3}
	\\
	& \left\lVert \mathbf{v}_{i} - \mathbf{v}_{i}^{\text{o}} \right\rVert \leq a_{\rm max}\tau,  \quad \forall i\in \mathcal{I},   \label{uav_con4}
	\\
	& d_{i,j} \geq d_{\rm min}, \quad \forall i,j\in \mathcal{I}, \forall i\neq j. \label{uav_con5}
\end{align}
\end{subequations}
\endgroup
Constraint \eqref{uav_con1} determines its current location based on its previous one $\mathbf{X}_{i}^{\text{o}}$ with respect to (w.r.t.) current AM-RIS velocity. Constraint \eqref{uav_con2} confines its flying area. Constraints \eqref{uav_con3} and \eqref{uav_con4} control the maximum AM-RIS flight movement velocity $v_{\rm max}$ and acceleration $a_{\rm max}$, respectively, where $\mathbf{v}_{i}^{\text{o}}$ indicates the previous velocity. Constraint \eqref{uav_con5} prevents the collision among AM-RISs.

We consider a switch-based AM-RIS, where each element is configured only by either energy harvesting mode (H-mode) or signal reflection mode (S-mode). The configuration of the $i$-th AM-RIS is defined as $\boldsymbol{\Theta}_i = \diag \left( \alpha_{i,1} \sqrt{\beta_{i,1}} e^{j\theta_{i,1}},...,\alpha_{i,M} \sqrt{\beta_{i,M}} e^{j\theta_{i,M}} \right)$, where notation of $\alpha_{i,m}\in\{0,1\}$ indicates operation mode, i.e., $\alpha_{i,m}=1$ indicates the S-mode, whereas $\alpha_{i,m}=0$ represents the H-mode. Moreover, $\beta_{i,m} =(0, \beta_{\rm max}]$ and $\theta_{i,m}\in [0, 2\pi)$ denote the AM-RIS amplitude and phase-shifts, respectively, with $\beta_{\rm max}$ defined as its maximum amplification gain.

\subsection{FA-Empowered Array}
The $n_a$-th element of the FA array response vector $\mathbf{a}_{{\rm FA}, a}\in \mathbb{C}^{N_a}$, where $a\in\{\rm T,R\}$ can be expressed as
\begin{align} \label{FAR}
	[\mathbf{a}_{{\rm FA}, a}]_{n_a} = e^{j \mathbf{k}_a \mathbf{x}_{n_a}^{\rm FA}}, 
\end{align}
where the wave vector is denoted as $\mathbf{k}_a = \frac{2\pi}{\lambda} \left[ 
\sin \psi_a  \sin \vartheta_a, 
\sin  \psi_a \cos \vartheta_a, 
\cos  \psi_a 
\right]$, $\lambda$ is the wavelength, and the position of each FA element is $\mathbf{x}_{n_a}^{{\rm FA}} = [x^{{\rm FA}}_{n_a}, y^{{\rm FA}}_{n_a}, z^{{\rm FA}}_{n_a} ]^{\mathcal{T}}$. Notations of $\psi_a$ and $\vartheta_a$ with $a\in\{t,r\}$ are azimuth and elevation angles-of-departure (AoD) ($a={\rm T}$)/angles-of-arrival (AoA) ($a={\rm R}$) of FA Tx/Rx at FD-BS, respectively. Due to the hardware limitation, FA element of FD-BS is geometrically confined by the following constraints \cite{my4}
\begingroup
\allowdisplaybreaks
\begin{subequations} \label{FA_con}
\begin{align}
	&  0 \leq x^{{\rm FA}}_{n_{\rm T}} \leq L_x, \quad 
	\forall n_{\rm T} \in \mathcal{N}_{\rm T}, \label{FA_con1}
	\\
	&  L_x \leq x^{{\rm FA}}_{n_{\rm R}} \leq 2 L_x, \quad 
	  \forall n_{\rm R}\in\mathcal{N}_{\rm R}, \label{FA_con2}
	\\
	&  0 \leq y^{{\rm FA}}_{n_a} \leq L_y, \quad 
      \forall n_a\in\mathcal{N}_a, \forall a\in\{\rm T,R\}, \label{FA_con3}
	\\
	& z^{{\rm FA}}_{n_a} = z^{\rm BS}, \quad
	  \forall n_a\in\mathcal{N}_a, \forall a\in\{\rm T,R\}, \label{FA_con4}\\
	& \left\lVert \mathbf{x}_{n_a}^{{\rm FA}} - \mathbf{x}_{n'_a}^{{\rm FA}} \right\rVert \geq d_{{\rm th},1}, \quad 
      \forall n_a, n'_a \in\mathcal{N}_a, \forall a\in\{\rm T,R\}, \label{FA_con5}\\
      & \left\lVert \mathbf{x}_{n_{\rm T}}^{{\rm FA}} - \mathbf{x}_{n_{\rm R}}^{{\rm FA}} \right\rVert \geq d_{{\rm th},2}, \quad 
      \forall n_{\rm T}\in\mathcal{N}_{\rm T}, \forall n_{\rm R}\in\mathcal{N}_{\rm R}. \label{FA_con6}
\end{align}
\end{subequations}
\endgroup
Constraints \eqref{FA_con1}--\eqref{FA_con3} constrain the deployable xy-region for each element at Tx/Rx, whereas \eqref{FA_con4} limits their identical height to that of the FD-BS. Moreover, constraints \eqref{FA_con5} and \eqref{FA_con6} impose minimum inter-element spacing requirements, where $d_{{\rm th},1}$ governs the intra-antenna spacing and $d_{{\rm th},2}$ governs the inter-array spacing, respectively.

\subsection{AM-RIS Channel Modeling}

The array response vector of AM-RIS can be expressed as $\mathbf{b}_{a,i}\in\mathbb{C}^{M}$:
\begin{align} \label{steer}
	& \mathbf{b}_{a,i} \!=\! [1, e^{j\frac{2\pi}{\lambda} d_{A} \sin \psi_{a,i} \sin \vartheta_{a,i} },..., e^{j\frac{2\pi}{\lambda}(M_x\!-\!1) d_{A} \sin \psi_{a,i} \sin \vartheta_{a,i} }]^{\mathcal{T}} \notag\\
	& \otimes [1, e^{j\frac{2\pi}{\lambda} d_{A} \sin \psi_{a,i} \cos \vartheta_{a,i} },..., e^{j\frac{2\pi}{\lambda} (M_y\!-\!1) d_{A} \sin \psi_{a,i} \cos \vartheta_{a,i} }]^{\mathcal{T}},
\end{align}
where $\{\psi_{a,i}, \vartheta_{a,i}\}$ indicate the azimuth and elevation angles of AM-RIS $i$, with $a\in\{\rm T,R\}$ as AoD and AoA, respectively. Notation of $d_A$ is the fixed inter-element spacing of the AM-RIS. The channel between the BS Tx and AM-RIS $i$ is modeled as Rician channel $\mathbf{D}_{{\rm T},i} \in \mathbb{C}^{M \times N_{\rm T}}$, given by \cite{mfris1}
\begin{align} \label{ric}
	\mathbf{D}_{{\rm T},i} \!=\! \sqrt{h_0 d_{i}^{-\kappa_0}} \left( \sqrt{\frac{\beta_0}{1\!+\!\beta_0}} \mathbf{D}^{\rm LoS}_{{\rm T},i}  \!+\! \sqrt{\frac{1}{1\!+\!\beta_0}} \mathbf{D}^{\rm NLoS}_{{\rm T},i} \right),
\end{align}
where $h_0$ is the reference pathloss at $1$ meter, $d_i=\lVert \mathbf{X}^{\rm BS} -  \mathbf{X}_{i} \rVert$ means the distance between the BS and AM-RIS $i$, and $\beta_0$ indicates the pathloss exponent. The LoS component combines the array vectors of Tx and AM-RIS, i.e., $\mathbf{D}^{\rm LoS}_{{\rm T},i}= \mathbf{b}_{{\rm R},i} \mathbf{a}_{{\rm FA},{\rm T}}^{\mathcal{H}}$. While, the NLoS path $\mathbf{D}^{\rm NLoS}_{{\rm T},i} $ is modeled as Rayleigh fading, following a complex Gaussian distribution with zero mean and unit variance. Similarly, we can define the channel from the AM-RIS $i$ to BS Rx as $\mathbf{D}_{{\rm R},i}\in \mathbb{C}^{N_{\rm R} \times M}$ following the same definitions as \eqref{ric} in conjunction with its LoS component defined as $\mathbf{D}^{\rm LoS}_{{\rm R},i} = \mathbf{a}_{{\rm FA}, {\rm R}} \mathbf{b}_{{\rm T},i}^{\mathcal{H}}$. We consider that no direct link is available from the BS to users, whereas the signals have be delivered via the AM-RIS. We consider the symmetric channel between the AM-RIS $i$ and user $k$, defined as $\mathbf{g}_{i,k} \in \mathbb{C}^{M}$:
\begin{align}
	\mathbf{g}_{i,k} \!=\! \sqrt{h_0 d_{i,k}^{-\kappa_0}} \left( \sqrt{\frac{\beta_0}{1\!+\!\beta_0}} \mathbf{g}^{\rm LoS}_{i,k}  \!+\! \sqrt{\frac{1}{1\!+\!\beta_0}} \mathbf{g}^{\rm NLoS}_{i,k} \right),
\end{align}
where $d_{i,k} = \lVert  \mathbf{X}_{i} -  \mathbf{X}_{k}^{\rm UE} \rVert$ is the distance between AM-RIS $i$ and user $k$. The LoS part is denoted as $\mathbf{g}^{\rm LoS}_{i,k} = \mathbf{b}_{a,i,k}, \forall a\in\{\rm T,R\}$ where $\mathbf{b}_{a,i,k}$ following \eqref{steer} stands for the steering vector for user $k$. Note that $\mathbf{g}^{\rm NLoS}_{i,k}$ follows Rayleigh fading. The channel between the AM-RIS $i$ and $j$ can be defined as $\mathbf{H}_{i,j}\in \mathbb{C}^{M\times M}$:
\begin{align}
	\mathbf{H}_{i,j} \!=\! \sqrt{h_0 \bar{d}_{i,j}^{-\kappa_0}} \left( \sqrt{\frac{\beta_0}{1\!+\!\beta_0}} \mathbf{H}^{\rm LoS}_{i,j}  \!+\! \sqrt{\frac{1}{1\!+\!\beta_0}} \mathbf{H}^{\rm NLoS}_{i,j} \right),
\end{align}
where $\bar{d}_{i,j}= \lVert \mathbf{X}_{i}- \mathbf{X}_{j} \rVert$ is the distance between two AM-RISs $i$ and $j$. The LoS path component is acquired as $\mathbf{H}^{\rm LoS}_{i,j} = \mathbf{b}_{{\rm T},j} \mathbf{b}_{{\rm R},i}^{\mathcal{H}}$, whereas NLoS portion $\mathbf{H}^{\rm NLoS}_{i,j}$ follows Rayleigh fading. We take into account the non-negligible inter-AM-RIS channels, as second-order reflections, such as the propagation path from the BS through AM-RIS 1 and AM-RIS 2 to the user, can still contribute to the overall channel response and affect system performance. Then the combined channel between the FD-BS to user $k$ can be obtained as $\mathbf{h}_{{\rm T},k}\in\mathbb{C}^{1 \times N_{\rm T}}, \forall k\in\mathcal{K}_{\rm T}$ and $\mathbf{h}_{{\rm R},k}\in\mathbb{C}^{N_{\rm R} \times 1}, \forall k\in\mathcal{K}_{\rm U}$:
\begingroup
\allowdisplaybreaks
\begin{align}
	\mathbf{h}_{{\rm T},k} &\!=\! \sum_{i\in\mathcal{I}} \mathbf{g}_{i,k}^{\mathcal{H}} \boldsymbol{\Theta}_i \mathbf{D}_{i,{\rm T}} \!+ \!
	\sum_{i\in\mathcal{I}} \sum_{j\in\mathcal{I}\backslash  i }
	\mathbf{g}_{j,k}^{\mathcal{H}} \boldsymbol{\Theta}_j \mathbf{H}_{i,j} \boldsymbol{\Theta}_i \mathbf{D}_{i,{\rm T}}, \label{ch1} \\
	\mathbf{h}_{{\rm R},k} &\!=\! \sum_{i\in\mathcal{I}}  \mathbf{D}_{i,{\rm R}} \boldsymbol{\Theta}_i \mathbf{g}_{i,k} \!+\! 
	\sum_{i\in\mathcal{I}} \sum_{j\in\mathcal{I}\backslash  i }
	\mathbf{D}_{j,{\rm R}} \boldsymbol{\Theta}_j \mathbf{H}_{i,j} \boldsymbol{\Theta}_i \mathbf{g}_{i,k}. \label{ch2}
\end{align}
\endgroup
Note that the SI channel from Tx to Rx at FD-BS is defined as $\mathbf{S} \in \mathbb{C}^{N_{\rm R} \times N_{\rm T}}$, whereas the inter-user interference channel from user $k\in\mathcal{K}_{\rm D}$ to $k'\in\mathcal{K}_{\rm U}$ is denoted as $h_{k,k'}$. The notation of $\mathbf{S}$ follows \eqref{ric} with large $\beta_0$ due to the deterministic dominant direct path. Note that FA does not reduce SI by signal processing, but reduces SI by geometry-aware adjustment decoupling the antenna positions. However, designing FA simultaneously aligning SI nulling from direct and AM-RIS-induced paths as well as enhancing the UL/DL signals becomes compellingly challenging. To elaborate further, the LoS probability for the flying AM-RIS is given by
\begin{align}
P_{\rm LoS}(\vartheta_{a,i}) = \frac{1}{1+ b_1 \cdot e^{ b_2 (\vartheta_{a,i}- b_1)}},
\end{align}
where $b_1$ and $b_2$ are environment parameters determined by the building density and height \cite{uav_prob}.

The UL power and the UL signal of user $k$ are denoted as $p_{k}$ and  $x_{{\rm U},k} $, respectively. While, the DL beamforming and the corresponding DL signal for user $k$ at Tx of the FD-BS are defined as $\mathbf{w}_{k}$ and $ x_{{\rm D},k} $, respectively. Then the UL signals received at BS Rx and the DL signals received at user $k$ can be respectively given by
\begingroup
\allowdisplaybreaks
\begin{align}
	& \mathbf{y}_{\rm U} = \sum_{k \in \mathcal{K}_{\rm U}} \underbrace{\mathbf{h}_{{\rm R},k} \sqrt{p_{k}} x_{{\rm U},k}}_{\text{UL Signal}} 
	+ \sum_{k \in \mathcal{K}_{\rm D}} \underbrace{ \bar{\mathbf{F}}_{i} \mathbf{w}_{k} x_{{\rm D},k} }_{\text{AM-RIS-Induced SI}} \notag \\
	& \qquad\qquad + \sum_{i \in \mathcal{I}} \underbrace{\mathbf{F}_{{\rm R},i} \mathbf{n}_{i}	}_{\text{AM-RIS Noise}}
	+ \mathbf{n}, \label{r_ul} \\
	& y_{{\rm D},k} = \underbrace{\mathbf{h}_{{\rm T},k} \mathbf{w}_k x_{{\rm D},k}}_{\text{DL Signal}} + \sum_{k'\in\mathcal{K}_{\rm D} \backslash k} \underbrace{\mathbf{h}_{{\rm T},k} \mathbf{w}_{k'} x_{{\rm D},k'}}_{\text{DL Interf.}} \notag \\
	& + \sum_{k'\in\mathcal{K}_{\rm U}} \underbrace{ \bar{f}_{k,k'} \sqrt{p_{k'}} x_{{\rm U},k'}}_{\text{AM-RIS-Induced UL  Interf.}} 
	+ \sum_{i \in \mathcal{I}} \underbrace{ \mathbf{f}_{i,k} \mathbf{n}_{i}	}_{\text{AM-RIS Noise}}	
	+ n_k , \label{r_dl}
\end{align}
\endgroup
where the AM-RIS-induced SI channel is denoted as $\bar{\mathbf{F}}_{i} = \sum_{i\in\mathcal{I}} \mathbf{F}_{{\rm R},i} \mathbf{D}_{{\rm T},i} + \mathbf{S}$ and channel related to AM-RIS-induced UL interference is attained as $ \bar{f}_{k,k'} = \sum_{i\in\mathcal{I}} \mathbf{f}_{i,k} \mathbf{g}_{i,k'} + h_{k,k'}$. Notation $\mathbf{F}_{{\rm R},i} = \mathbf{D}_{{\rm R},i} \boldsymbol{\Theta}_{i} + \sum_{j\in\mathcal{I}\backslash i} \mathbf{D}_{{\rm R},j} \boldsymbol{\Theta}_{j} \mathbf{H}_{i,j} \boldsymbol{\Theta}_{i} \in \mathbb{C}^{N_{\rm R} \times M}$ indicates the effective channel from AM-RIS $i$ to the Rx at FD-BS. Moreover, the effective from AM-RIS $i$ to DL user $k$ is expressed as $\mathbf{f}_{{\rm R},i} = \mathbf{g}_{i,k}^{\mathcal{H}} \boldsymbol{\Theta}_{i} + \sum_{j\in\mathcal{I}\backslash i} \mathbf{g}_{j,k}^{\mathcal{H}} \boldsymbol{\Theta}_{j} \mathbf{H}_{i,j} \boldsymbol{\Theta}_{i} \in \mathbb{C}^{1 \times M}$. Notations of $\mathbf{n}$, $\mathbf{n}_{i}$ and $n_k$ denote the noise induced by the FD-BS, AM-RIS $i$ and user $k$, respectively. It can be inferred from \eqref{r_ul} and \eqref{r_dl} that the FA array affects the BS Tx-Rx channel of $\mathbf{D}_{i,{\rm T/R}}$ and $\mathbf{S}$ by adjusting the antenna position, capable of directly suppressing SI. In parallel, the AM-RIS provides electromagnetic control over the cascaded BS-RIS-user links via $\boldsymbol{\Theta}_i$, enabling the generation of destructive SI components along the Tx-RIS-Rx path to further mitigate SI. Since the AM-RIS is also responsible for alleviating inter UL-DL user interference in addition to SI, the FA introduces additional spatial degrees of freedom that enhances SI suppression capability. Then we can attain the signal-to-interference-plus-noise ratio (SINR) of UL/DL users respectively as:
\begingroup
\allowdisplaybreaks
\begin{align}
	& \gamma^{\rm UL}_{k} = \frac{ p_{k} \lVert \mathbf{h}_{{\rm R},k} \rVert^2}{
	\sum\limits_{k \in \mathcal{K}_{\rm D}} \left\lVert  \bar{\mathbf{F}}_{i} \mathbf{w}_{k} \right\rVert^2 + \sum\limits_{i\in\mathcal{I}} \bar{\sigma}^2 \lVert \mathbf{F}_{{\rm R},i} \rVert_F^2 + \sigma^2
},
\\
	&\gamma^{\rm DL}_{k} =\notag
	\\
	& \frac{ | \mathbf{h}_{{\rm T},k} \mathbf{w}_k |^2}{ \smashoperator[r]{\sum\limits_{k'\in \mathcal{K}_{\rm D}  \backslash k}} |\mathbf{h}_{{\rm T},k} \mathbf{w}_{k'} |^2 + \smashoperator[r]{\sum\limits_{k' \in \mathcal{K}_{\rm U}}} p_{k'}  | \bar{f}_{k,k'}|^2 + \sum\limits_{i\in\mathcal{I}} \bar{\sigma}^2
 \lVert \mathbf{f}_{i,k} \rVert^2 \!+\! \sigma_k^2,
	}
\end{align}
\endgroup
where $\bar{\sigma}^2$, $\sigma^2$, and $\sigma^2_{k}$ are noise power of AM-RIS, BS, and DL user $k$, respectively. $\lVert \cdot \rVert_F^2$ indicates the Frobenius norm. The corresponding user rates are given by
\begin{align}\label{rate}
	R^{\rm DL}_{k} = \log \Big( 1 + \gamma^{\rm DL}_{k} \Big),  \quad
	R^{\rm UL}_{k} = \log \Big( 1 + \gamma^{\rm UL}_{k} \Big).
\end{align}


\subsection{Power Dissipation Model}

\subsubsection{Power Model of AAV}
We consider the rotary-wing AAV for AM-RIS which requires propulsion power to support its movement as well as its aerial hovering. The mechanical power consumption of AM-RIS $i$ is modeled as \cite{uav_move1}
\begingroup
\allowdisplaybreaks
	\begin{align}
	P_{i}^{\rm AAV} &= P_{\rm BP} \left( 1+ \frac{3 \left\lVert \mathbf{v}_{i} \right\rVert^2}{ \Omega_{ B}^2 \Omega_{ R}^2}\right)
	+ \frac{1}{2}\zeta_{ D} \zeta_{ A} \zeta_{ S} \zeta_{ R} \left\lVert \mathbf{v}_{i} \right\rVert^2 \notag\\
	 &+ P_{\rm IP} \left[  \left(  1+\frac{\left\lVert \mathbf{v}_{i} \right\rVert^4}{4 v_{ R}^4} \right)^{\frac{1}{2}} -  \frac{\left\lVert \mathbf{v}_{i} \right\rVert^2}{2 v_{ R}^2} \right]^{\frac{1}{2}},
	\end{align}
	\endgroup
where $\Omega_{ B}$ represents the blade angular velocity (rad/s) and $\Omega_{ R}$ indicates rotor radius (m). Notations of $\zeta_{ D}$, $\zeta_{ A}$, $\zeta_{ S}$, and $\zeta_{ R}$ denote the fuselage drag ratio, air density, rotor solidity, and rotor disk area of the AAV, respectively. The average rotor induced velocity is given by $v_{ R}$. Moreover, the blade profile power in hovering mode is expressed by
	\begin{align}
	P_{\rm BP} = \frac{1}{8} \zeta_{ P} \zeta_{ A} \zeta_{ S} \zeta_{ R} \Omega_B^3 \Omega_R^3,
	\end{align}
	where $\zeta_{P}\geq 0$ is a profile drag coefficient. The induced power in hovering mode is given by
	\begin{align}
	P_{\rm IP} = (1+\zeta_C) W_l^{\frac{3}{2}} \left( 2 \zeta_A \zeta_R \right)^{\frac{-1}{2}},
	\end{align}
	where $\zeta_C \geq 0$ is an incremental correction factor in the induced power, whereas $W_i$ is the weight of AM-RIS $i$ with the unit of Newton. We further notice that the weight includes both the AAV as well as the MF-RIS, i.e., $W_i = W_i^{\rm AAV} + W_i^{\rm MRIS}$. To elaborate a little further, $W_i^{\rm MRIS} \propto M$ is proportional to the number of AM-RIS elements since more elements need larger manufacturing size of the metasurface. Subsequently, the hovering operation power can be derived as $P_{{\rm H},i}^{\rm AAV} = P_{\rm BP} + P_{\rm IP}$ by setting $\left\lVert \mathbf{v}_{i} \right\rVert = 0$.

\subsubsection{Power Model of MF-RIS}
	In FD networks supported by AM-RISs, the harvested energy can come from the Tx at BS, UL users as well as the induced noise from other AM-RISs. Firstly, we define the EH coefficient matrix as $\mathbf{E}_{i,m} = {\rm diag}( [0,...,0, 1-\alpha_{i,m},0,...,0])$, where the $m$-th element is with the factor $1-\alpha_{i,m}$. Then, the radio frequency (RF) power received at the $m$-th element of the $i$-th AM-RIS is given by
\begin{align} \label{P_rf}
	P^{\rm RF}_{i,m} = \mathbb{E} 
	\left\lbrace 
	\left\lVert 
	\mathbf{E}_{i,m}
	\left( 
 	\smashoperator[r]{\sum_{k\in \mathcal{K}_{\rm D}}} \bar{\mathbf{D}}_{i} \mathbf{w}_{k} 
	+ \smashoperator[r]{\sum_{k\in\mathcal{K}_{\rm U}}} \bar{\mathbf{g}}_{i} \sqrt{p_k}  
	+ \mathbf{n}_{i} 
	\right) 
	\right\rVert^2 
	\right\rbrace,
\end{align}
where
$\bar{\mathbf{D}}_{i} = \mathbf{D}_i + \sum_{j\in\mathcal{I}\backslash i} \mathbf{H}_{j,i} \boldsymbol{\Theta}_{j} \mathbf{D}_{{\rm T},j}$ and $\bar{\mathbf{g}}_{i} = \mathbf{g}_{i,k} + \sum_{j\in\mathcal{I}\backslash i} \mathbf{H}_{j,i} \boldsymbol{\Theta}_{j} \mathbf{g}_{j,k}$. The expectation in \eqref{P_rf} is computed w.r.t. $\mathbf{n}_{i}$ since they cannot be precisely estimated. Therefore, the power harvested at the $m$-th element of the $i$-th AM-RIS follow the non-linear circuital property, given by \cite{mfris1}
\begin{align} \label{EH_model}
	P_{i,m}^{\rm H} = \frac{\Gamma_{i,m} - Z_1 Z_2}{1-Z_2}, 
\end{align}
where $\Gamma_{i,m} = Z_1/\left( 1+e^{-c_1 \left(P^{\rm RF}_{i,m} - c_2 \right)} \right)$ is a characteristic function w.r.t. the input signal power at RF $P_{i,m}^{\rm RF}$. The constant $Z_1 \geq 0$ indicates the maximum available harvested power. The constant $Z_2 = \frac{1}{1+e^{c_1 c_2}}$ with $c_1 > 0$ and $c_2 > 0$ captures the joint effects of energy harvesting circuit sensitivity limitations and leakage. Then the total AM-RIS power consumption excluding the mechanical part is modeled as \cite{RIS_power1}
\begin{align} \label{Pris}
	& P^{\rm MRIS}_{i} = \max \left[ 0, \sum_{m\in \mathcal{M}} \alpha_{i,m}  \cdot (P_{\rm ph} + P_{\rm am}) \right. \notag \\
	& \left. 
	+ \left( M - \sum_{m\in \mathcal{M}} \alpha_{i,m} \right) \cdot P_{\rm hc} + \varsigma_{\rm PA} P^{\rm out}_{i} - \sum_{m\in \mathcal{M}} P^{\rm H}_{i,m} \right],
\end{align}
where $P_{\rm ph}$, $P_{\rm am}$ and $P_{\rm hc}$ represent power consumed by each phase-shift, amplitude control, and EH conversion circuit, respectively. Notation $\varsigma_{\rm PA}$ is the inverse of the power amplifier efficiency. In \eqref{Pris}, the output power of the $i$-th AM-RIS is given by
\begin{align}
	P^{\rm out}_{i} & \!=\!  \left\lVert \boldsymbol{\Theta}_{i} \smashoperator[r]{\sum_{k\in \mathcal{K}_{\rm D}}} \bar{\mathbf{D}}_{i} \mathbf{w}_{k}  \right\rVert^2
	\!+\! \left\lVert \boldsymbol{\Theta}_{i} \smashoperator[r]{\sum_{k\in\mathcal{K}_{\rm U}}} \bar{\mathbf{g}}_{i} \sqrt{p_k}  \right\rVert^2 
	\!+\! 
	\bar{\sigma}^2
 \left\lVert \boldsymbol{\Theta}_{i} \right\rVert^2_{F} .
\end{align}
Accordingly, the total power consumption\textsuperscript{\ref{note1}}\footnotetext[1]{AM-RIS flight power may dominate the EE performance when more AM-RISs are deployed, i.e., the mechanical power becomes more non-negligible, accounting for about $55\%$-$88\%$ of the total power consumption. In contrast, when only a few AM-RISs are involved, the communication power can be dominant. Moreover, when the AM-RIS operates in the hovering mode, generally incurring approximately twice the power consumption compared to the moving mode at the optimal speed, and serves scenarios such as arenas or specific hotspots, the communication power consumption can be negligible. \label{note1}} of MF-RIS-AAV $i$ can be obtained as $P_{i} = P^{\rm MRIS}_{i} + P^{\rm AAV}_{i}$. The EE\textsuperscript{\ref{note2}}\footnotetext[2]{In this work, EE mainly focuses on mechanical flight of AAVs and communication power. The computational energy of the algorithm agents is implicit and considered fixed as the proposed framework can assume lightweight onboard training and inference, whose power is negligible. However, the model and problem should be redesigned if considering the dynamic computing power, algorithmic models and different capability under massive data amount, which is left as future potentials. \label{note2}} can be expressed by
\begin{align}
EE \triangleq \frac{ \sum_{k \in \mathcal{K}_{\rm D}} R^{\rm DL}_{k_{\rm D}} 
    + \sum_{k \in \mathcal{K}_{\rm U}} R^{\rm UL}_{k_{\rm U}} }
    { \sum_{i\in \mathcal{I}} P_{i}
    + \sum_{k \in \mathcal{K}_{\rm D}} \left\lVert \mathbf{w}_{k} \right\rVert^2 
    + \sum_{k \in \mathcal{K}_{\rm U}} p_k }.
\end{align}

\subsection{Problem Formulation} 
We formulate an EE maximization problem w.r.t. the arguments of DL beamforming $\mathbf{w}_{k}$ and UL power $p_{k}$ of UL user $k$, AM-RIS configuration $\boldsymbol{\Theta}_i$, FA position $\mathbf{x}_{n_a}^{\rm FA}$ of Tx/Rx at FD-BS, and deployment of AM-RIS $\mathbf{X}_i$, represented by
\begingroup
\allowdisplaybreaks
 \begin{subequations} \label{problem_T}
\begin{align}
    &\mathop{\max} \limits_{\substack{  {\mathbf{w}_{k}, p_{k}, \mathbf{x}_{n_a}^{\rm FA}, \boldsymbol{\Theta}_i}, \mathbf{X}_i}} \quad  
     EE \label{obj} \\
     \text{s.t. } & \quad \boldsymbol{\Theta}_{i} \in \mathcal{F}_{\Theta}, \quad \forall i \in \mathcal{I}, \label{con1} \\
     &\quad \sum_{k \in \mathcal{K}_{\rm D}} \left\lVert \mathbf{w}_{k} \right\rVert^2 \leq P_{\rm BS}, \label{con2} \\
     &\quad p_{k} \leq P_{\rm UL}, \quad \forall k \in \mathcal{K}_{\rm U}, \label{con3} \\
     &\quad P_{i} \leq P_{\rm max}, \quad \forall i \in \mathcal{I}, \label{con4}\\
     &\quad R^{\rm DL}_{k} \geq R_{\rm th}^{\rm DL}, \quad \forall k\in \mathcal{K}_{\rm D}, \label{con5} \\
     &\quad R^{\rm UL}_{k} \geq R_{\rm th}^{\rm UL}, \quad \forall k\in \mathcal{K}_{\rm U}, \label{con6}\\
     &\quad \mathbf{X}_{i} \in \mathcal{F}_{X}, \quad \forall i \in \mathcal{I} \label{con7},\\
     &\quad \mathbf{x}_{n_a}^{\rm FA} \in \mathcal{F}_{\rm FA}, \quad \forall n_a \in \mathcal{N}_a, \forall a\in\{\rm T,R\}. \label{con8}
\end{align}
 \end{subequations}
  \endgroup
Constraint \eqref{con1} indicates the AM-RIS configuration constraints in the feasible domain $\mathcal{F}_{\Theta}=\{\alpha_{i,m}\in\{0,1\}, \beta_{i,m} \in (0, \beta_{\rm max}], \theta_{i,m} \in [0, 2\pi) | \forall m \in \mathcal{M} \}$. Constraints \eqref{con2} and \eqref{con3} exhibits the DL and UL power constrained by $P_{\rm BS}$ and $P_{\rm UL}$, respectively. In \eqref{con4}, power consumption of AM-RIS should be lower than the given threshold of $P_{\rm max}$. Constraints \eqref{con5} and \eqref{con6} guarantee the minimum per user rate requirement of DL/UL to be higher than $R^{\rm DL}_{\rm th}$ and $R^{\rm UL}_{\rm th}$, respectively. Constraint \eqref{con7} represents the restriction of AM-RIS deployment as $\mathcal{F}_{X}=\{\eqref{uav_con1}$--$\eqref{uav_con5}\}$.
Constraint \eqref{con8} limits the position adjustment of FA at Tx/Rx as $\mathcal{F}_{\rm FA}=\{\eqref{FA_con1}$--$\eqref{FA_con6}\}$. We can observe that it is difficult to directly solve the non-convex and non-linear problem \eqref{problem_T}. To address the challenge, we propose an deep reinforcement learning based scheme to obtain the solution.

\section{Proposed SOHRL Algorithm}\label{sec_alg}

We have conceived a SOHRL framework for solving problem \eqref{problem_T}, which consists of a multi-agent architecture, where each node acts as an agent equipped with a pair of DQN and PPO to determine its own policy consisting of multiple continuous and discrete variables. Furthermore, attention mechanism is adopted to focus on informative states, whereas self-optimization technique is designed to reduce the manual trials of hyperparameters, thereby accelerating the overall learning speed and enhancing training efficiency.

\begin{figure*}[!t]
	\centering
	\includegraphics[width=6.6in]{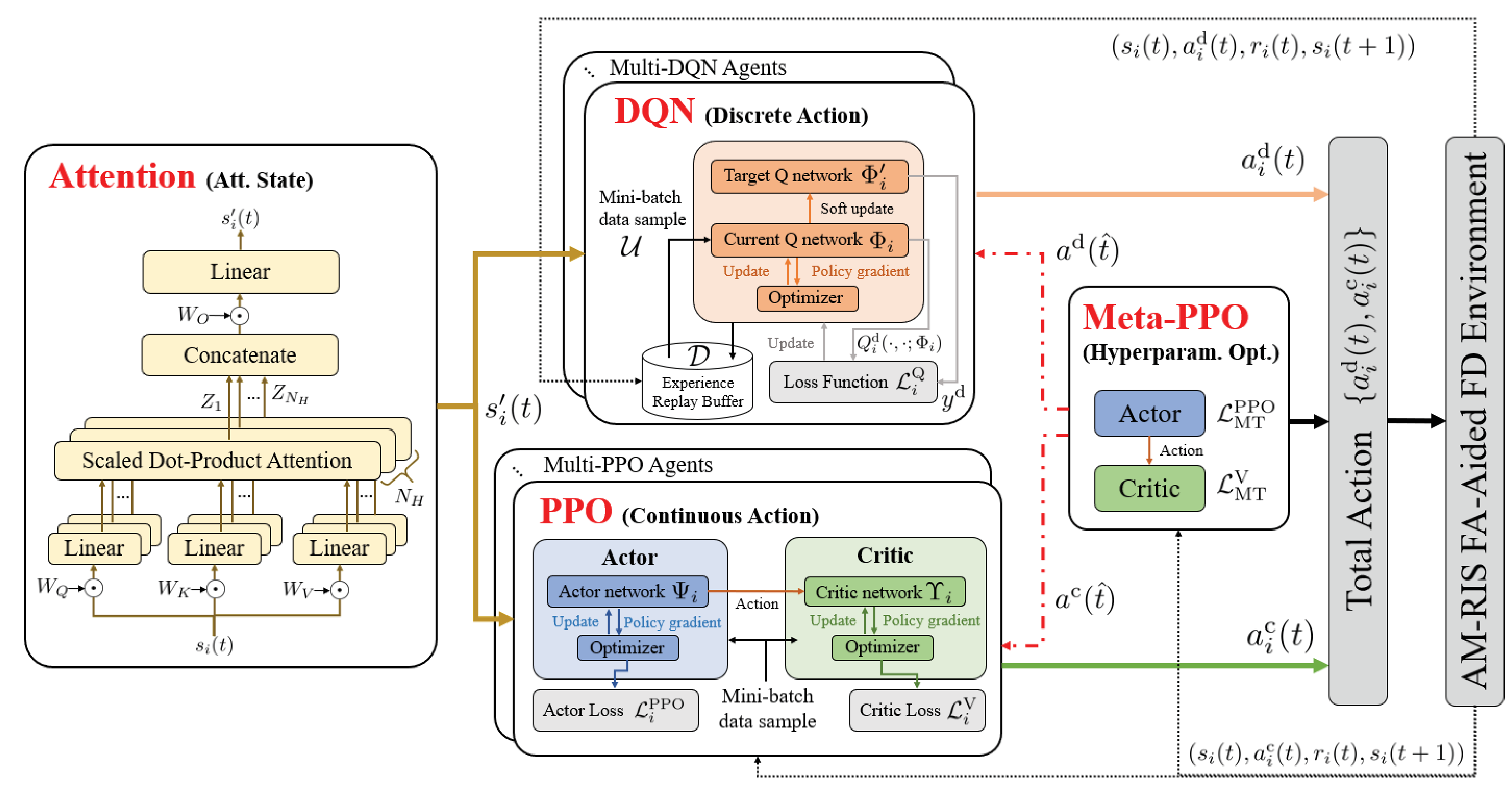}
	\caption{The proposed SOHRL algorithm integrating MADQN and MAPPO with attention-driven states and self-optimization scheme.}
	\label{proposed}
\end{figure*}

\subsection{Multi-Agent Hybrid DRL}
 
A typical DRL framework is employed, which comprises the state set $\mathcal{S}$, action set $\mathcal{A}$ and reward set $\mathcal{R}$. We utilize a multi-agent system to improve the training  and decision-making efficiency, as shown in Fig. \ref{proposed}. In the designed framework, an agent standing for either the FD-BS or an AM-RIS interacts with the wireless environment by performing \textit{actions}, attaining the resulting \textit{rewards}, and updating their corresponding \textit{states}, which are defined as follows.

\subsubsection{Attention-Driven State Representation}

The total state set is denoted as individual agent state as $\mathcal{S} = \{\mathcal{S}_1, \mathcal{S}_2, ..., \mathcal{S}_{I}, \mathcal{S}_{I+1} | \forall i \in \bar{\mathcal{I}} \}$, where $\bar{\mathcal{I}} = \mathcal{I}\cup \{I+1\}$, and index $I+1$ refers to the FD-BS. Each node has the state $\mathcal{S}_{i} = \{s_{i}(1), s_{i}(2), ..., s_{i}(T)\}$, where $T$ is the total timeslots of the system. The individual state of the $i$-th AM-RIS at time $t$ is designed as the combined channel related to AM-RIS $i$ to user $k$, i.e., $s_{i}(t) = \{ \mathfrak{R}\{\mathbf{g}^{\rm DL}_{k}(t)\}, \mathfrak{I}\{\mathbf{g}^{\rm DL}_{k}(t)\}, \mathfrak{R}\{\mathbf{g}^{\rm UL}_{k'}(t)\}, \mathfrak{I}\{\mathbf{g}^{\rm UL}_{k'}(t)\} | \forall k \in \mathcal{K}_{\rm D}, \forall k' \in \mathcal{K}_{\rm U} \}, \forall i\in\mathcal{I}$, where $\mathbf{g}^{\rm DL}_{k}(t) = \mathbf{g}_{i,k}^{\mathcal{H}}(t)\cdot \boldsymbol{\Theta}_i(t) \cdot \mathbf{D}_{i,{\rm T}}(t)$ and $\mathbf{g}^{\rm UL}_{k'}(t) = \mathbf{g}_{i,k'}^{\mathcal{H}}(t) \cdot \boldsymbol{\Theta}_i (t) \cdot \mathbf{D}_{i,{\rm T}}(t)$. Note that $\mathfrak{R}\{\cdot\}/\mathfrak{I}\{\cdot\}$ extracts the real/imaginary parts of a complex variable. Similarly, the individual state of the FD-BS can be given by $s_{I+1}(t) = \{ \mathfrak{R}\{ \mathbf{h}_{{\rm T},k}(t) \}, \mathfrak{I}\{ \mathbf{h}_{{\rm T},k}(t) \}, \mathfrak{R}\{\mathbf{h}_{{\rm R},k'}(t)\}, \mathfrak{I}\{\mathbf{h}_{{\rm R},k'}(t)\} | \forall k \in \mathcal{K}_{\rm D}, \forall k' \in \mathcal{K}_{\rm U}\}$. To prevent overhead of excessive channel information exchange, we consider that only the channel related to the node itself can be measured. 

Nevertheless, considering the state with equal weights is impractical as different channels potentially provide different levels of important information. Therefore, we incorporate the attention mechanism \cite{atten} to provide unequal weights focused on each state element. Accordingly, instead of relying solely on raw channel observations $s_i(t)$, each agent firstly processes its state through a multi-head self-attention mechanism, associated with the linearly projected \textit{query}, \textit{key} and \textit{value}, i.e.,
\begin{align}
 	Y_Q = W_{Q} s_{i}(t),  \quad
 	Y_K = W_{K} s_{i}(t), \quad 
 	Y_V = W_{V} s_{i}(t),
\end{align}
where $W_{Q/K/V}$ are trainable projection parameters. Then the attention output is computed using the scaled dot-product as
\begin{align}\label{att}
	Z = \mathrm{Att}(Y_Q, Y_K, Y_V) = \mathrm{Softmax} \left( \frac{Y_Q Y_K^{\mathcal{T}}}{\sqrt{d_K}} \right) \cdot Y_V,
\end{align}
where $\mathrm{Att}(\cdot)$ is the attention operation and $\mathrm{Softmax}(\mathbf{x}) = \frac{e^{x_i}}{\sum_{j} e^{x_j}}$ is the softmax function. Notation $d_K$ is the dimension of the parameter of $Y_K$, i.e., the feature size of the key space of $W_K$. To prevent overfitting of a single attention model, multi-head attention is conceived at left of Fig. \ref{proposed} to generate the attention-driven states $s'_i(t)$ by concatenating the obtained attention heads, given by
\begin{align}
	s'_{i}(t) \!=\! \mathrm{MH}(Y_Q, Y_K, Y_V) \!=\! \mathrm{ConCat}(Z_1, Z_2, ..., \mathrm{Z}_{N_H}) \cdot W_O,
\end{align}
where $\mathrm{MH}(\cdot)$ is multi-head operation, $\mathrm{ConCat}(\cdot)$ indicates concatenation operation, and $Z_{n_H} = \mathrm{Att}(\cdot)$ is the attention-based input based on \eqref{att}. Notation of $N_H$ means the number of input heads, whilst $W_O$ is the trainable parameter of multi-head output. To elaborate a little further, the query $Y_Q$ encodes the agent's objective, guiding attention toward relevant features of channel conditions. The key matrix $Y_K$ serves to signify the importance of different input features, whereas the value $Y_V$ carries the corresponding contextual content used for appropriate decision-making.

\subsubsection{Continuous-Discrete Actions of Each Agent}

The total action set is partitioned into continuous and discrete portions as $ \mathcal{A}^p = \{\mathcal{A}^p_{1}, \mathcal{A}^p_{2}, \ldots, \mathcal{A}^p_{i} | \forall i\in\bar{\mathcal{I}} \}, \forall p \in \mathcal{B} = \{\text{c},\text{d}\}$, where 'c' and 'd' indicate continuous and discrete variables, respectively. The action set of agent $i$ is denoted as $\mathcal{A}^p_{i} = \{a^p_{i}(1), a^p_{i}(2), ..., a^p_{i}(T)\} $. Note that the AM-RIS configurations of amplitude and phase-shift $\{\beta_{i,m}(t), \theta_{i,m}(t)\}$, FD-BS beamforming $\mathbf{w}_{k}(t)$, uplink user power $p_{k'}(t)$ are continuous variables. Meanwhile, EH ratio of AM-RIS $\alpha_{{i},m}(t)$, AAV deployment position $\mathbf{X}_{i}(t)$ and FA position $\mathbf{x}_{n_a}^{\rm FA}$ are considered discrete variables, where the geometric positions are quantized to facilitate simpler hardware implementations. A multi-port FA architecture is adopted \cite{fa2,fa3,fa4}, where multiple antenna ports are selectively activated from a set of predefined high-resolution FA positions for transmission and reception at FD-BS. To avoid the prohibitively large action space induced by binary on-off port selection, we instead adopt a movement-based mechanism, where the positions of each FA element are updated sequentially. Therefore, the dimension of the FA action space is reduced to $5\cdot (N_{\rm T}+N_{\rm R})$, where $5$ corresponds to the discrete movement options at each resolution grid, i.e., motion along the $\pm$x-axis, $\pm$y-axis, or remaining stationary is performed for each Tx and Rx antenna. Accordingly, the hybrid action $a^p_{i}(t)$ of AM-RIS agent $i$ at time $t$ is defined as $ a^{\text{c}}_{i}(t) = \{ \beta_{i,m}(t), \theta_{i,m}(t) | \forall m\in \mathcal{M}\}$ for continuous part and $a^{\text{d}}_{i}(t) = \{ \alpha_{{i},m}(t), \mathbf{X}_{i}(t) | \forall m\in \mathcal{M}\} , \forall i\in\mathcal{I}$ for discrete case. The hybrid action at $a^p_{I+1}(t)$ of the FD-BS at time $t$ is denoted as $ a^{\text{c}}_{I+1}(t) = \{ \mathfrak{R}\{\mathbf{w}_{k}(t)\}, \mathfrak{I}\{\mathbf{w}_{k}(t)\}, p_{k'}(t) | \forall k\in \mathcal{K}_{\rm D}, \forall k' \in \mathcal{K}_{\rm U}\}$ for continuous part and $a^{\text{d}}_{i}(t) = \{ \mathbf{x}_{n_a}^{\rm FA} | \forall n_a \in \mathcal{N}_a, \forall a\in\{\rm T, R\}\}$ for discrete case.


\subsubsection{System Reward} 

The reward set is defined as $\mathcal{R} = \{\mathcal{R}_{1}, \mathcal{R}_2, ..., \mathcal{R}_{I}, \mathcal{R}_{I+1}| \forall i\in\bar{\mathcal{I}} \}$, where the individual reward is given by $\mathcal{R}_{i} = \{r_{i}(1), r_{i}(2), ..., r_{i}(T)\}$. The reward $r_{i}(t)$ is obtained and commonly shared after performing all actions, which is designed as EE with the penalty terms, given by
\begin{align}
	r_{i}(t) = EE(t) - \sum_{c=1}^{6} \rho_{c} C_{c},
\end{align}
where $\rho_{c} \geq 0$ indicates the weights of each constraint penalty $C_{c}$ corresponding to the constraints in problem \eqref{problem_T}, which is designed as 
\begingroup
\allowdisplaybreaks
\begin{subequations}
\begin{align}
	C_{1} &\!=\! \sum_{k\in\mathcal{K}_{\rm D}} \left( R_{\rm th}^{\rm DL} - R_{k}^{\rm DL} \right), \
	C_{2} \!=\! \sum_{k\in\mathcal{K}_{\rm U}} \left( R_{\rm th}^{\rm UL} - R_{k}^{\rm UL} \right) , \\
	C_{3} & \!=\! \sum_{i \in \mathcal{I}} P_i \!-\! P_{\rm max}, \
	C_{4} \!=\!  \sum_{i=1}^{I-1} \sum_{j=i+1}^{I} \left( d_{\rm min} \!-\! d_{i,j} \right), \\
	C_{5} & \!=\! \sum_{a\in\{\rm T, R \}}\sum_{n_a=1}^{N_a-1}  \sum_{n'_a=n_a+1}^{N_a} \left( d_{{\rm th},1} \!-\!  \lVert \mathbf{x}_{n_a}^{{\rm FA}} \!-\! \mathbf{x}_{n'_a}^{{\rm FA}} \rVert \right), \\
	C_{6} &= 
\sum_{n_{\rm T} \in \mathcal{N}_{\rm T}} \sum_{n_{\rm R} \in \mathcal{N}_{\rm R}}
\left( d_{{\rm th},2} -  \lVert \mathbf{x}_{n_{\rm T}}^{{\rm FA}} - \mathbf{x}_{n_{\rm R}}^{{\rm FA}} \rVert \right).
\end{align}
\end{subequations}
\endgroup
It is worth noting that the remaining boundary constraints are inherently satisfied during the action generation process by using the function of $x = \min( \max(x_{\rm min}, x), x_{\rm max})$, corresponding to its upper/lower bounds of $x_{\rm max/min}$. We also note that transmit beamforming should be normalized if exceeding its power budget, i.e., $\mathbf{w}_{k}(t) = \frac{\mathbf{w}_{k}(t)}{ \sum_{k\in\mathcal{K}_{\rm D}} \lVert \mathbf{w}_{k}(t) \rVert^2} \cdot P_{\rm BS}$ if $\sum_{k \in \mathcal{K}_{\rm D}} \left\lVert \mathbf{w}_{k} \right\rVert^2 \geq P_{\rm BS}$. We further note that a shared global reward $r_i(t)$ across agents may induce non-stationarity and credit-assignment ambiguity. In the proposed framework, stability is achieved by a combination structural, temporal and algorithmic design: Although actions are taken by multiple agents, the environment is driven by a single global system objective of penalized EE in a cooperative manner. Hence, all agents aim for optimizing the same objective, which removes reward inconsistency across agents and avoids conflicting gradients causing competitive instability. Even with the common reward, each agent action impacts the reward through a well-defined pathway, i.e., the reward gradient w.r.t. each action remains informative but has weak inter-agent coupling, significantly reducing credit ambiguity. Furthermore, the target network within each agent will periodically update the neural network model from the current network based on the soft update for potential stabilization, as detailed in later paragraphs.


As shown in Fig. \ref{proposed}, the proposed framework integrates multi-agent DQN (MADQN) and multi-agent PPO (MAPPO) to tackle discrete and continuous actions, respectively. Note that PPO and DQN networks share the same input state $s'_{i}(t)$, ensuring consistent observations and collaborative learning process. A \textit{current-target} network structure is employed in DQN for stabilizing the training process, which is defined as $\Phi_i$ and $\Phi'_i$, respectively. The current network $\Phi_i$ also termed as Q-network is responsible for jointly selecting and evaluating discrete actions. While, the target Q-network of DQN $\Phi'_i$ intends to stabilize the learning process by periodically copying the models from their current networks. The discrete actions are generated by exploration-exploitation policy in current networks to prevent local solutions: (1) Each DQN agent selects the best action by either its Q-function as $ a_{i}^{\text{d}}(t) = \argmax_{a^{\text{d}}_{i} \in \mathcal{A}^{\text{d}}_{i} } {Q}_{i}^{\text{d}} (s_{i}(t),a^{\text{d}}_{i}; \Phi_i)$ if $\varsigma \geq \varsigma_{\rm th}$ or random selection otherwise, where $\varsigma$ is a randomly generated value and $\varsigma_{\rm th}$ is the exploration threshold. Note that ${Q}_i^{\text{d}}$ is the current Q network; In PPO, the \textit{policy} network $\Psi_i$ takes care of action generation, whilst the \textit{value} network $\Upsilon_i$ aims for evaluating the action. Each PPO agent determines its action by Gaussian distribution $ a_{i}^{\text{c}}(t) = \pi_i (s_{i}(t); \Psi_i) = \mathcal{N}( \mu_{\Psi_i}(s_i(t)), \Lambda_{\Psi_i}(s_i(t)) )$, where $\mu_{\Psi_i}(\cdot)$ and $\Lambda_{\Psi_i}(\cdot)$ are mean and variance. The reward $r_{i}(t)$ will be obtained and the state will be updated as $s_{i}(t+1) = s_{i}(t)$ after performing the total action $a_i(t) = \{a_{i}^{\text{d}}(t), a_{i}^{\text{c}}(t)\}$.

The memory replay buffer $\mathcal{D}$ with a size of $|\mathcal{D}|$ is constructed to store the trajectory for MADQN training, i.e., historical experiences, with the tuple of $\Xi_{i}(t) = (s_{i}(t), a^{\text{d}}_{i}(t), r_{i}(t), s_{i}(t+1))$. We randomly select $U \leq |\mathcal{D}|$ samples from $\mathcal{D}$ for the neural network training, where the data sample set is defined as $\mathcal{U}=\{1,2,...,U\}$. Based on data at time instant $u \in  \mathcal{U}$, the corresponding values of MADQN is acquired as
\begin{align}
    y^{\text{d}} = r_{i}(u) + \gamma_{\text{d}} \cdot \max_{a_{i}^{\text{d}} \in\mathcal{A}^{\text{d}}_{i} } {Q}_i^{\text{d}} \left(s_{i}(u + 1), a_{i}^{\text{d}}; \Phi'_i \right),
\end{align}
where $\gamma_{\text{d}} \in [0, 1] $ is discount factor indicating the importance of future rewards of MADQN. To minimize the estimation error between target and the current networks, the loss function of MADQN is formulated as \cite{dqn}
\begin{align}
    \mathcal{L}^{Q}_i =  \mathbbm{E}_{t, \mathcal{U}} \left[ \Big( y^{\text{d}} - Q_{i}^{\text{d}} \left(s_{i}(t), a_{i}^{\text{d}}(t) ; \Phi_i \right) \Big)^2 \right].\label{dqnloss}
\end{align}
The stochastic gradient descent (SGD) is employed to update the model weights of the current network in MADQN, i.e., $\Phi_i \leftarrow \Phi_i - \eta_{\text{d}} \cdot \nabla_{\Phi_i} \mathcal{L}^{Q}$, where $\eta_{\text{d}}$ is the learning rate. The gradient of MADQN in \eqref{dqnloss} is calculated as $\nabla_{\Phi_i} \mathcal{L}^{Q} \approx \mathbbm{E} [ 2 (y^{\text{d}} - Q_{i}^{\text{d}} (s_{i}(t), a_{i}^{\text{d}}(t); \Phi_i) ) \cdot \nabla_{\Phi_i} Q_{i}^{\text{d}} (s_{i}(t), a_{i}^{\text{d}}(t); \Phi_i) ]$. To elaborate further, the target network will periodically update the neural network model per $T_{up}$ steps from the current network based on the soft update \cite{my7} as
\begin{align} \label{softt}
   \Phi'_i & \leftarrow \tau_{Q} \Phi_i + (1 - \tau_{Q}) \Phi'_i,
\end{align}
where $ 0\leq \tau_{Q}\leq 1 $ is a positive constant indicating the portion of parameters contributed by the current network.

In MAPPO, an \textit{advantage} function is adopted to measure how much better or worse taking action in certain state \cite{ppo}, i.e., 
\begin{align} \label{advan}
	& A_i(t) = Q^{\text{c}}_i(s_i(t), a^{\text{c}}_i(t))  -  V_{i} \left(s_{i}(t); \Upsilon_i \right) \notag \\
	& \approx r_{i}(t) + \gamma_{\text{c}} \cdot V_{i} \left(s_{i}(t+1); \Upsilon_i \right)	 -  V_{i} \left(s_{i}(t); \Upsilon_i \right),
\end{align}
where $Q^{\text{c}}_i$ is the Q-function of MAPPO. Notation $V_{i} \left(s_{i}(t); \Upsilon_i \right)$ represents the value function of MAPPO in conjunction with the evaluation neural network $\Upsilon_i$ given the state $s_i(t)$. The discount factor of $ \gamma_{\text{c}} \in [0, 1] $ indicates the importance of future rewards of MAPPO. Then a surrogate objective is employed via a clipped objective, given by
\begin{align} \label{clip}
\mathcal{L}^{\rm PPO}_i  = \mathbb{E} \left[ \min \left( \varrho(\Psi_i) A_i(t), G_i(t) A_i(t) \right) \right],
\end{align}
where $\varrho(\Psi_i) = \frac{\pi_i(s_i(t);\Psi_i)}{\pi_i(s_i(t);\Psi_{i,{\rm old}})}$ indicates the probability ratio of current policy to the previous policy, and $G_i(t)$ is the clipping function restricting the value to the interval $[1-\epsilon, 1+\epsilon]$, represented by
\begin{align}
	G_i(t) &= {\rm clip}( \varrho(\Psi_i) , 1-\epsilon, 1+\epsilon) \notag \\
	& = 
\begin{cases}
1-\epsilon, & \text{ if } \varrho(\Psi_i) < 1-\epsilon, \\
\varrho(\Psi_i),  &\text{ if } 1-\epsilon \leq \varrho(\Psi_i) \leq 1+\epsilon, \\
1+\epsilon,  &\text{ if } \varrho(\Psi_i) > 1+\epsilon.
\end{cases}
\end{align}
It is worth mentioning that the loss function of \eqref{clip} performs two critical tasks, i.e.,  prevention of $\varrho(\Psi_i)$ from increasing policy probabilities too aggressively when $A_i(t) > 0$ and avoidance from decreasing them too drastically when $A_i(t) < 0$. Moreover, the value function loss is defined as
\begin{align} \label{valoss}
	\mathcal{L}^{\rm V}_i = \mathbb{E}_t \left[ \left( V_i(s_i(t); \Upsilon_i) - \hat{R}_{i}(t) \right)^2 \right],
\end{align}
where $ \hat{R}_{i}(t) = r_i(t) +\gamma_{\text{c}} \cdot V_i(s_i(t+1); \Upsilon_i)$ is the temporal difference return. The SGD is employed to update the model weights of the MAPPO, i.e., $\Psi_i \leftarrow \Psi_1 - \eta_{\text{c},1} \cdot \nabla_{\Psi_i} \mathcal{L}_{i}^{\rm PPO}$ and $\Upsilon_i \leftarrow \Upsilon_i - \eta_{\text{c},2} \cdot \nabla_{\Upsilon_i} \mathcal{L}_{i}^{\rm V}$, where $\eta_{\text{c},1}$ and $\eta_{\text{c},2}$ are learning rates. The corresponding gradient w.r.t. $\Psi_i$ is derived as $\nabla_{\Psi_i} \mathcal{L}^{\text{PPO}}_i \approx
\mathbb{E}_t \left[
\nabla_{\Psi_i} \log \pi_i(a_i(t) | s_i(t); \Psi_i) \cdot \tilde{g}_i(t)
\right]$,
where
\begin{align*}
\tilde{g}_i(t) =
\begin{cases}
\varrho(\Psi_i) \cdot A_i(t), & \text{if } \varrho(\Psi_i)  A_i(t) \leq G_i(t)  A_i(t), \\
G_i(t) \cdot A_i(t), & \text{otherwise}.
\end{cases}
\end{align*}
Also, the gradient w.r.t. $\Upsilon_i$ is attained as $\nabla_{\Upsilon_i} \mathcal{L}^{V}_i =
\mathbb{E}_t [
\left( V_i(s_i(t); \Upsilon_i) - \hat{R}_i(t) \right)
\cdot \left( \nabla_{\Upsilon_i} V_i(s_i(t); \Upsilon_i)
- \gamma_c \cdot \nabla_{\Upsilon_i} V_i(s_i(t+1); \Upsilon_i) \right)
]$.
To elaborate a little further, we note that unlike MADQN, MAPPO is an on-policy method, learning from fresh trajectories collected using the current policy $\pi_i(s_i(t); \Psi_i)$.

\subsection{Self-Optimization of Hyperparameters}

In any DRL techniques, the selection of hyperparameters, such as learning rate, discount factor, and exploration ratio, plays a pivotal role in determining the stability, convergence speed, and the system performance. However, static or manually tuned hyperparameters may become suboptimal in dynamic multi-agent environments. To circumvent this, a self-optimized hyperparameter scheme in SOHRL is proposed, which adaptively tunes the hyperparameters during training. Inspired by the PPO framework, we treat the hyperparameter adjustment process as a separate learning task. A dedicated meta-agent is trained to observe the current state of hybrid DRL, decide hyperparameter adjustments, and then obtain the corresponding reward based on the learning quality and system performance. We note that the meta-agent operating in the outer loop of the overall learning framework evolves on a longer time scale $\hat{t}$ compared to the hybrid DRL in the inner loop $t$, i.e., $\hat{t} = \varpi \cdot t$, where $\varpi \geq 1$ is a positive integer. This separation is essential, as frequent updates at the meta-level potentially lead to unstable policy behavior. The state, action and reward of the meta-agent is designed as follows.
\subsubsection{Meta-State}
	The meta-state space $\mathcal{S}_{\rm MT}$  comprises the current hyperparameter values. At time $\tau$, the hyperparameter state vector is defined as $s_{\rm MT}(\hat{t}) = \{ s^{\text{d}}(\hat{t}), s^{\text{c}}(\hat{t}) \}$, where the meta-state for MADQN is 
$s^{\text{d}}(\hat{t})= \{ \gamma_{\text{d}}(\hat{t}),  \eta_{\text{d}}(\hat{t}),  \varsigma_{\rm th}(\hat{t}), \tau_Q(\hat{t}) \}$ and that for MAPPO is $s^{\text{c}}(\hat{t}) = \{\gamma_{\text{c}}(\hat{t}), \eta_{\text{c},1}(\hat{t}), \eta_{\text{c},2}(\hat{t}), \epsilon(\hat{t})\}$.

\subsubsection{Meta-Action}
The meta-action taken by the meta-agent at time $\hat{t}$ is a set of adjustments to the current hyperparameter values, denoted as $a_{\rm MT}(\hat{t}) = \{ a^{\text{d}}(\hat{t}), a^{\text{c}}(\hat{t}) \}$, where the meta-action for MADQN is
$a^{\text{d}}(\hat{t})= \{ \Delta\gamma_{\text{d}}(\hat{t}),  \Delta\eta_{\text{d}}(\hat{t}),  \Delta\varsigma_{\rm th}(\hat{t}), \Delta\tau_Q(\hat{t}) \}$ and that for MAPPO is $a^{\text{c}}(\hat{t}) = \{ \Delta\gamma_{\text{c}}(\hat{t}), \Delta\eta_{\text{c},1}(\hat{t}), \Delta\eta_{\text{c},2}(\hat{t}), \Delta\epsilon(\hat{t})\}$. Then the hyperparmeters to be set is based on 
\begin{align} \label{hyp}
	\xi(\hat{t}) \leftarrow \max \left( 0, \xi(\hat{t}) +\Delta\xi(\hat{t}) \right), 
\end{align}
where $\xi \in \mathcal{G} = \{ \gamma_{\text{d}},  \eta_{\text{d}},  \varsigma_{\rm th}, \tau_Q, \gamma_{\text{c}}, \eta_{\text{c},1}, \eta_{\text{c},2}, \epsilon\}$ denotes the set of all possible hyperparameters. These are updated incrementally, with each update applied to the corresponding hyperparameter within the hybrid DRL framework. Note that the first four elements of $\mathcal{G}$ indicate the discount factor, learning rate, exploration rate, and soft-update coefficient in MADQN, respectively. While, the last four elements of $\mathcal{G}$ denote the shared discount factor, learning rates of actor and critic as well as the clipping threshold in MAPPO, respectively. We further note that the meta-action is executed once every $\varpi$, meaning that $\xi$ remains fixed throughout $\varpi$ iterations of the inner-loop hybrid DRL process. Note that the tunable range of those hyperparameters are defined as $\xi \in [\xi_{\rm min}, \xi_{\rm max}]$ where $\xi_{\rm min/max}$ indicates the minimum and maximum values of the hyperparameters, respectively. We further notice that linear decay is not adopted in MADQN, as it tends to drive the learning process toward local optima in later iterations, which should be addressed through intelligent hyperparameter optimization.

\subsubsection{Meta-Reward}
The meta-reward function is designed to reflect both EE objective and the learning stability. Specifically, the reward at time $\hat{t}$ is designed as
\begin{align} \label{eq:reward}
	r_{\rm MT}(\hat{t}) =  \frac{1}{\varpi} \sum_{t'=0}^{\varpi-1} EE(t + t')  - \sum_{c=1}^{|\mathcal{G}|} \lambda_c \nu_c,
\end{align}
where the first term measures averaged EE performance after performing the hyperparameters for $\varpi$ iterations at inner loop. The second term is the penalty for learning instability, where $\nu_c$ is the penalty term associated with $\lambda_c$ defined as the corresponding weight. The penalty discourages the excessive deviation from the averaged hyperparameter settings, which is obtained as
\begin{align}
	\nu_c = \left( \frac{ \xi(\hat{t}) - \frac{1}{T_{\xi}} \sum_{t'=1}^{T_{\xi}} \xi(\hat{t}-t')  }{ \Delta \xi(\hat{t}) + \varepsilon_{\xi}} \right)^2,
\end{align}
where $T_{\xi}$ is the window size, and $\varepsilon_{\xi}$ is a small constant avoiding explosion of the value. We softly regularize deviations from the reference averaged hyperparameters to ensure stable exploration, while allowing certain adaptation. The meta-agent employs a single-agent PPO at outer loop by incorporating the similar models from \eqref{advan}--\eqref{valoss}, where the pertinent loss of policy and value function is given by
\begin{align}
	\mathcal{L}^{\rm PPO}_{\rm MT}  &= \mathbb{E} \left[ \min \left( \varrho(\Psi_{\rm MT}) A_{\rm MT}(t), G_{\rm MT}(t) A_{\rm MT}(t) \right) \right], \label{meta1}\\
	\mathcal{L}^{\rm V}_{\rm MT} &= \mathbb{E}_t \left[ \left( V_{\rm MT}(s_{\rm MT}(t); \Upsilon_{\rm MT}) - \hat{R}_{{\rm MT}}(t) \right)^2 \right], \label{meta2}
\end{align}
where $\Psi_{\rm MT}$ and $\Upsilon_{\rm MT}$ are the neural network weights of policy network and of value network, respectively. The remaining parameters follow the similar definitions to those in MAPPO at inner loop. The concrete algorithm of SOHRL is elaborated in Algorithm \ref{alg}. We notice that the proposed multi-agent learning process of SOHRL is distributed across agents, where each agent independently optimizes its own action $\mathcal{A}^{p}_i$, as described in Lines 6-16. These processes correspond to the "Attention, DQN, and PPO" modules illustrated in Fig. \ref{proposed}. In contrast, the meta-agent responsible for tuning all hyperparameters operates at the BS to enhance overall network stability, as indicated in Lines 22-25. This part is aligned with the "Meta-PPO" module as illustrated in Fig. \ref{proposed}. In this work, the multi-agent coordination in the proposed SOHRL scheme in Algorithm \ref{alg} assumes that control signaling is carried over a dedicated low-overhead control channel that is already required for network synchronization and AM-RIS configurations. The exchanged information consists only of the scalar rewards rather than raw channel state information, so the payload size is much smaller compared to the user data traffic. Consequently, its impact on the achievable rate and EE is negligible relative to the dominant data traffic transmission and mechanical energy terms considered in the proposed system model.

\begin{algorithm}[!t]
\caption{Proposed SOHRL Algorithm}
\small
\SetAlgoLined
\DontPrintSemicolon
\label{alg}
\begin{algorithmic}[1]
\STATE Initialize environment, policy network and value network of MADQN, MAPPO, and meta-PPO
\STATE Initialize all hyperparameters $\xi\in\mathcal{G}$
\STATE Set iteration count of outer/inner loop $\hat{t}=t=1$
\REPEAT [Outer Loop]
	\REPEAT [Inner Loop]	
		\FOR{Agent $i\in \bar{\mathcal{I}}$}
			\STATE MADQN $i$ selects discrete actions $a^{\text{d}}_i(t)$ based on $\varsigma$-greedy
			\STATE MAPPO $i$ select continuous actions $a^{\text{c}}_i(t)$ based on policy network $\pi_i(s_i(t); \Psi_i)$
			\STATE Perform joint action $a_i(t)=\{ a^{\text{d}}_i(t), a^{\text{c}}_i(t) \}$
		\ENDFOR
		
		\FOR{Agent $i\in \bar{\mathcal{I}}$}
			\STATE Obtain reward $r_i(t)$ and update the state $s_i(t+1)$
			\STATE Store trajectory $\Xi_{i}(t)$ for MADQN training
			\STATE Update agent $i$'s MQDQN model $\Phi_i$ based on \eqref{dqnloss} \\ and \eqref{softt} using sampled data in $\mathcal{D}$
			\STATE Update agent $i$'s MAPPO models $\Psi_i$ in \eqref{clip} and $\Upsilon_i$ \\ in \eqref{valoss} using on-policy data generated by $\pi_{i}$
		\ENDFOR
		
		\STATE $t \leftarrow t+1$
		\IF{${\rm mod}(t-1, \varpi)=0$}
			\STATE Break and perform hyperparameter update
		\ENDIF 
	\UNTIL End of inner loop
	\STATE Meta-agent selects meta-actions $a_{\rm MT}(\hat{t})$ based on policy network $\pi_{\rm MT}(s_{\rm MT}(\hat{t}); \Psi_{\rm MT})$
	\STATE Set the hyperparameters based on \eqref{hyp} to inner loop
	\STATE Obtain meta-reward $r_{\rm MT}(\hat{t})$ and update state $s_{\rm MT}(\hat{t}+1)$
	\STATE Update meta-model based on \eqref{meta1} and \eqref{meta2}
	\STATE $\hat{t} \leftarrow \hat{t} +1$
\UNTIL End of system time $T$	
\end{algorithmic}
\end{algorithm}

\section{Numerical Results} \label{sec_sim}

\begin{table}[!t]
\small
\setstretch{1.15}
\caption{Simulation parameters} \label{param}
\centering
\begin{tabular}{|p{2.1cm}|p{5cm}|}
\hline
\textbf{Parameter} & \textbf{Value} \\ \hline
Network parameters & 
$I=4$, 
$M=32$, 
$K_{\rm D}=K_{\rm U}=4$, 
$N_{\rm T}=N_{\rm R}=16$, 
$h_0 = -20$ dB,
$\kappa_0 = 2.2$, 
$\beta_0  = 3$ dB, 
$P_{\rm BS} = 40$ W,
$P_{\max} = 100$ W,
$P_{\rm UL} = 1$ W,
$\sigma_{k}^2 = \bar{\sigma}^2 = \bar{\sigma}_k^2 = -80$ dBm,
$R_{\rm th}^{\rm DL}=R_{\rm th}^{\rm UL}=1$ bps/Hz,
\\ \hline
AAV parameters \cite{uav_move1,uav_prob} &
$W_i^{\rm AAV} = 20$ Newton,
$P_{\rm BP} = 79$ W, 
$P_{\rm IP} = 88$ W,
$z_{i}\in[50,500]$ m, 
$d_{\rm min}=6$ m,
$v_{\rm max}=15$ m/s,
$a_{\rm max}=1$ m/s$^2$,
$\Omega_B = 300$ rad/s, 
$\Omega_R = 0.4$ m, 
$v_R= 4$ m/s,
$\zeta_D= 0.3$, 
$\zeta_A= 1.225$ kg/m$^3$, 
$\zeta_S= 0.05$, 
$\zeta_R= 0.5$ m$^2$, 
$\zeta_C= 0.1$,
$b_1=12.08$,
$b_2=0.11$,
$d_{{\rm th},1} = 0.05$ m,
$d_{{\rm th},2} = 0.1$ m.
\\ \hline
MF-RIS parameters \cite{mfris1}&
$W_{i}^{\rm MRIS} = 10$ Newton, 
$P_{\rm ph} = 5$ mW, 
$P_{\rm hc} = 50$ mW, 
$P_{\rm am} = 100$ mW,
$\beta_{\max} = 3$,
$\varsigma_{\rm PA} = 1.1$,
$Z_1 = 24$ mW,
$c_1 = 150$, 
$c_2 = 0.014$ 
\\ \hline
DRL parameters &
$\varpi = 5$,
$ |\mathcal{D}| = 128$,
$ U = 64$,
$\varsigma_{\rm th}\in [0.05, 0.4]$, 
$\gamma_{\text{d}} \in [0.1, 0.99]$, 
$\eta_{\text{d}} \in [10^{-4}, 10^{-2}]$, 
$\tau_Q \in [0.7, 0.95]$, 
$\gamma_{\text{c}} \in [0.1, 0.99]$,  
$\epsilon \in [0.1, 0.3]$, 
$\eta_{\text{c},1} \in [10^{-4}, 10^{-2}]$, 
$\eta_{\text{c},2} \in [10^{-4}, 10^{-2}]$, 
$\rho_1=\rho_2=\rho_3 = 5\times 10^{-4}$, $\rho_4=\rho_5=\rho_6 = 3\times 10^{-5}$,
$\lambda_c = 3 \times 10^{-4}$, $\varepsilon_{\xi}= 10^{-6}$,
$N_H=4$,
$T_{\xi}=5$,
$d_K=64$,
$T_{up}=10$
\\ \hline
\end{tabular}
\label{table:example}
\end{table}

\begin{figure}[!t]
\centering
\includegraphics[width=3in]{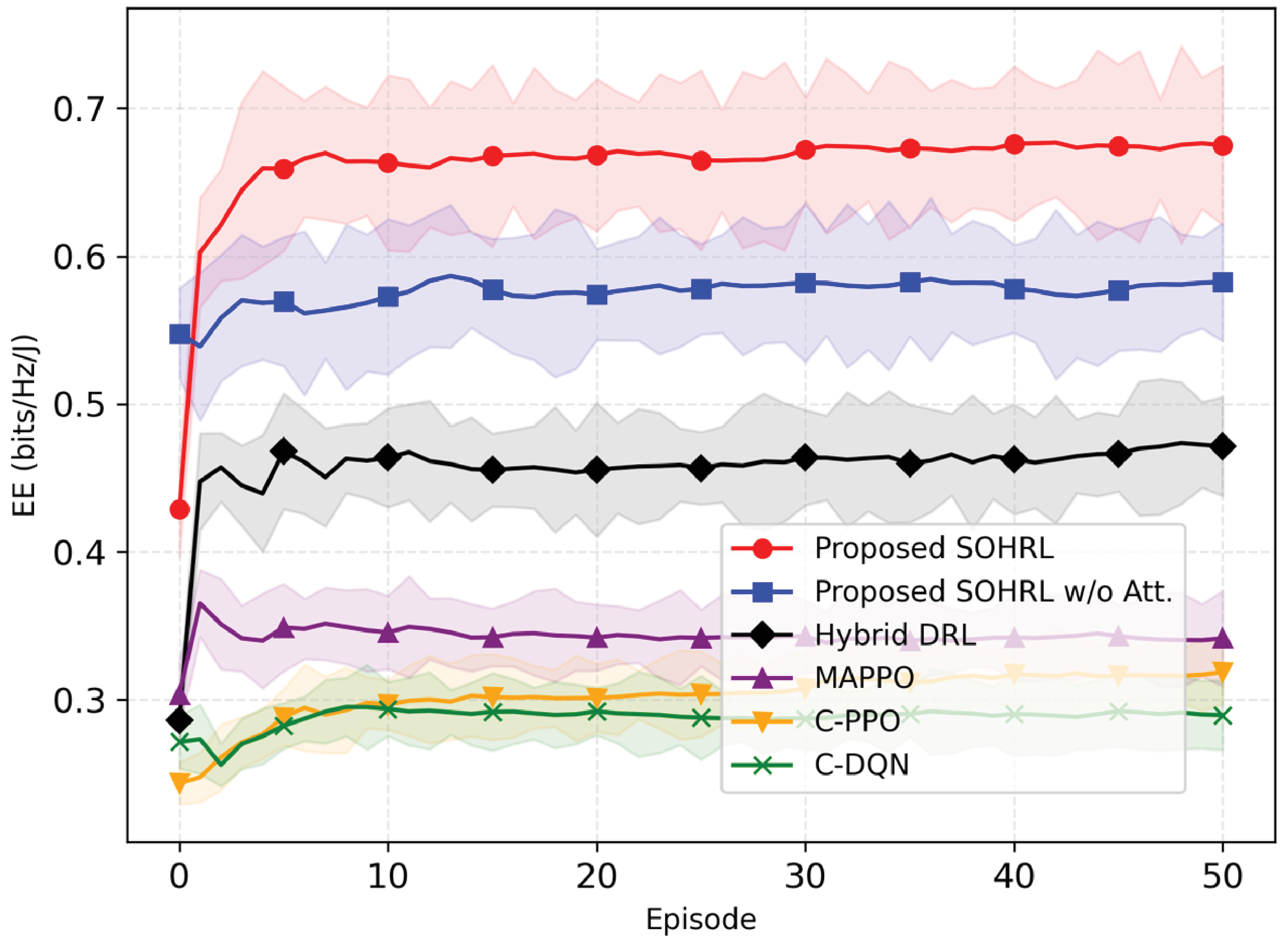}
\caption{Convergence of EE performance of proposed SOHRL scheme compared to the benchmarks of SOHRL without attention-drive state, hybrid DRL, MAPPO, centralized PPO and DQN methods.}
\label{fig:ee_comparison}
\end{figure}

We evaluate the performance of the proposed architecture of AM-RIS for the FA-aided FD-BS empowered by the conceived SOHRL algorithm. The serving area is set to $500 \times 500$ m$^2$, whereas the operating frequency is set to $3.5$ GHz. The pertinent parameters related to AAVs \cite{uav_move1,uav_prob}, MF-RISs \cite{mfris1} and DRL are listed in Table~\ref{param}. Fig.~\ref{fig:ee_comparison} presents the convergence of EE performance across different optimization strategies, including the proposed SOHRL scheme, SOHRL w/o attention-drive states (SOHRL w/o Att.), hybrid DRL \cite{HyRL_StarRIS}, MAPPO \cite{mappo}, centralized PPO (C-PPO) \cite{ppo} and centralized DQN \cite{dqn}. Note that hybrid DRL adopts a single neural network of DQN for dealing with discrete variables and deep deterministic policy gradient for continuous policy decisions. Moreover, C-PPO and C-DQN employ only one single neural network for all parameter decisions. We can observe that the proposed SOHRL scheme achieves the highest EE performance among all the other benchmarks, benefited from its hybrid DQN-PPO architecture, attention-driven states as well as hyperparameter optimizations. Without the attention mechanism, the input state information with equal importance weight leads to a degraded EE performance by around $15\%$. A lower EE performance of hybrid DRL is observed compared to SOHRL w/o Att. due to absence of hyperparameter optimization. For the learning policy of MAPPO, quantization errors from the continuous actor output variables approximated to the discrete actions result in a comparably lower EE performance. Moreover, without the processing of multi-agent systems, the benchmark of C-PPO tends to fall in the local optimum solution due to high-dimensional search spacing of possible action combinations. Owing to all discretized parameters and without the aid of multi-agents, the C-DQN performs the lowest EE performance among all methods. The observations above emphasize the necessary designs of SOHRL in hybrid continuous-discrete learning architecture, unequal importance weight from attention mechanism, and the self-optimization of hyperparameters.

\begin{figure}[!t]
    \centering
    \includegraphics[width=3in]{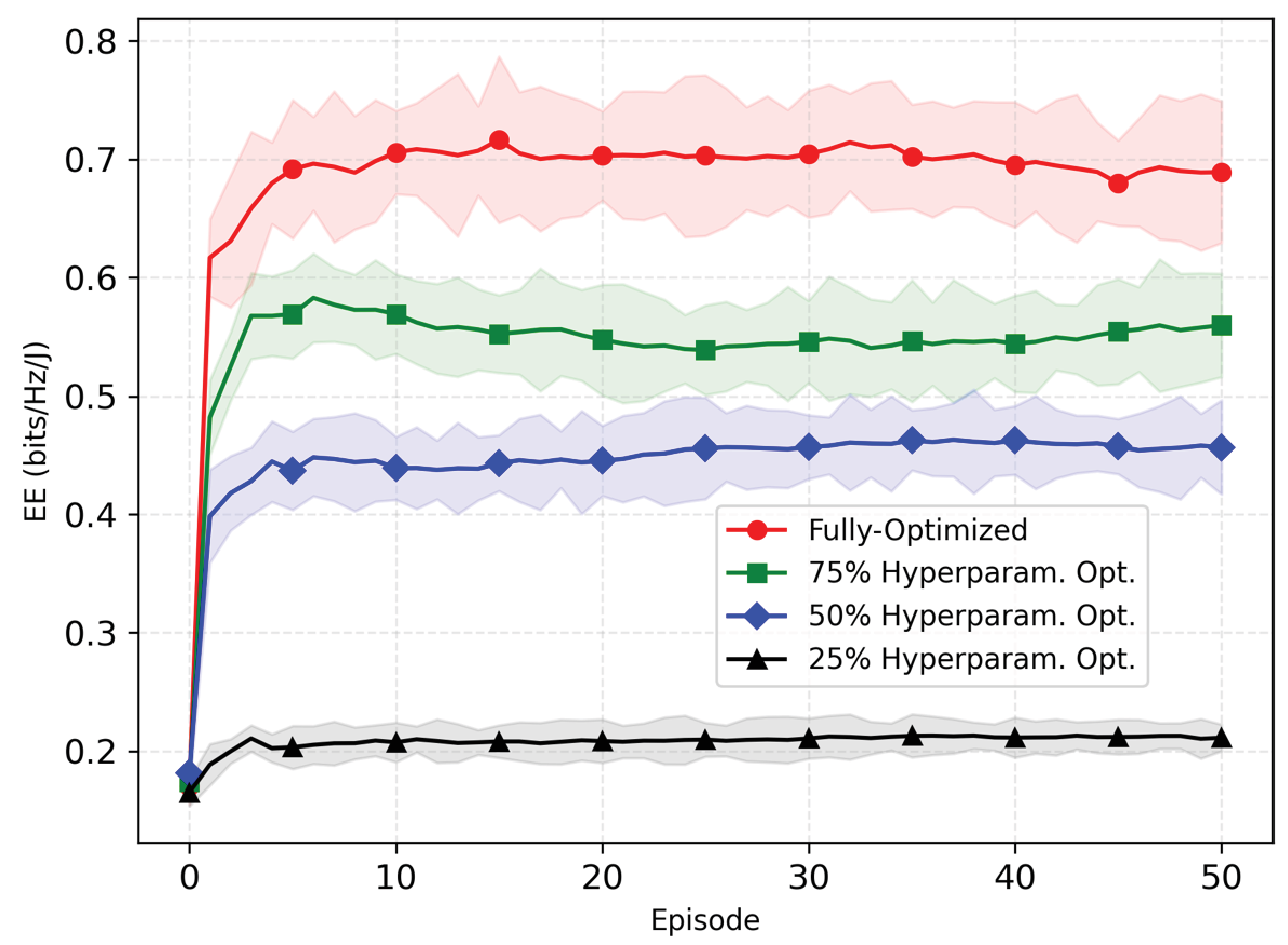}
    \caption{Convergence of EE performance of SOHRL with different portions of hyperparameters to be optimized.}
    \label{fig7}
\end{figure}

Fig.~\ref{fig7} illustrates the convergence of EE performance under different portions of hyperparameter optimization, i.e., the fully-optimized setting and partially optimized hyperprameters of $75\%$, $50\%$, and $25\%$. Note that we randomly select partial hyperparameters to be optimized in SOHRL. We can observe that the fully-optimized scheme achieves the highest EE performance, converging around $0.7$ bits/Hz/J, which highlights the importance of joint hyperparameter tuning in deep learning-based optimization frameworks. In contrast, when only $75\%$ of hyperparameters are optimized, the EE converges to approximately $0.55$ bits/Hz/J, showing an EE degradation of nearly $22\%$. Such drop becomes more significant when the optimization ratio is reduced further. When $50\%$ of hyperparameters are optimized, the EE stabilizes around $0.45$ bits/Hz/J, whereas EE degrades sharply to about $0.2$ bits/Hz/J for the case with $25\%$ hyperparameters optimized, indicating that insufficient hyperparameter adaptation in dynamic wireless environments severely deteriorates the system EE performance. Overall, the results emphasize the necessity of full hyperparameter optimization to maximize EE performance.

\begin{figure}[t]
    \centering
    \includegraphics[width=3in]{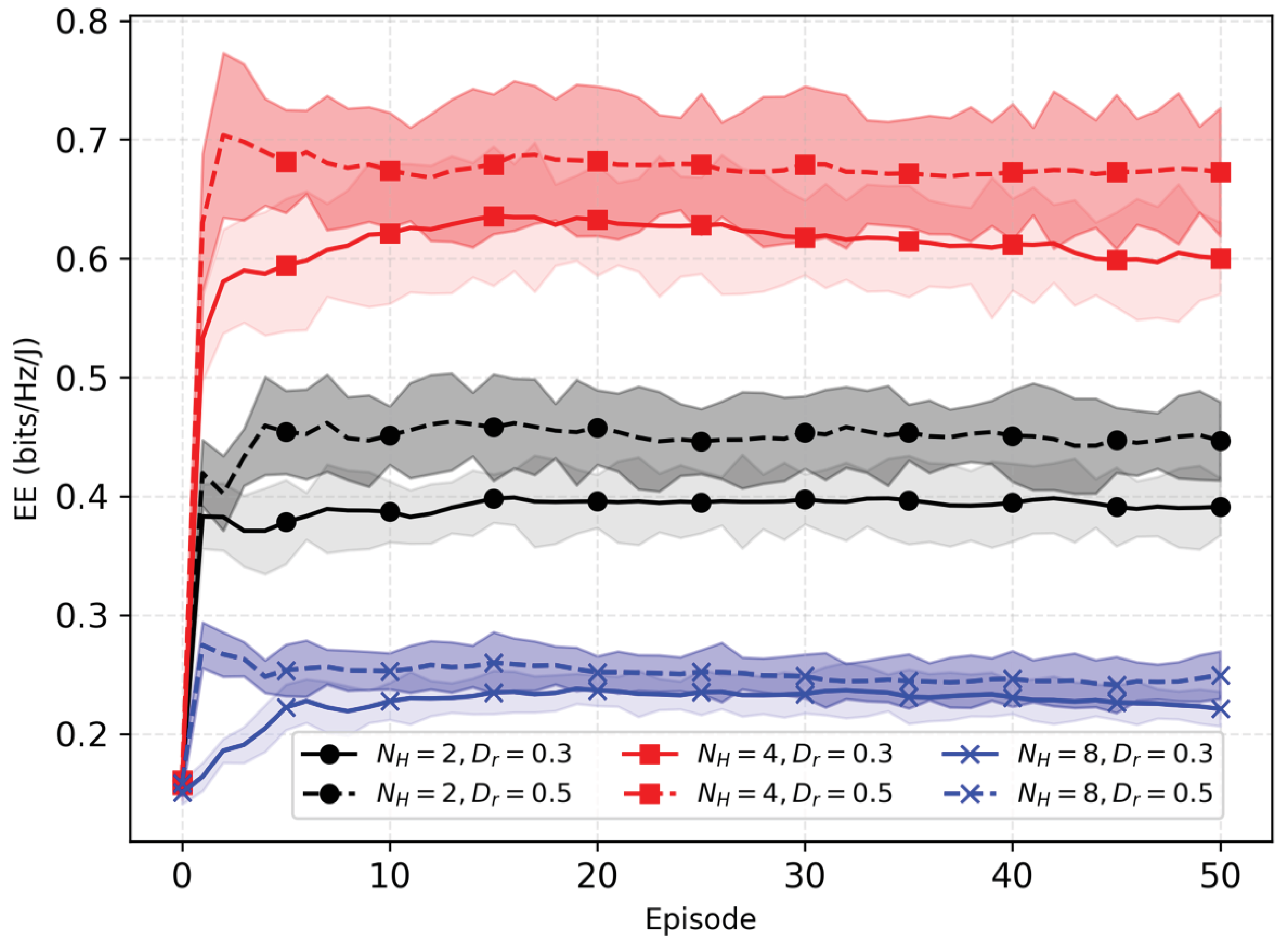}
    \caption{Convergence of EE performance of SOHRL with different numbers of the multi-head attention $N_H$ and dropout rates $D_r$ in deep neural networks.}
    \label{fig8}
\end{figure}

Fig.~\ref{fig8} demonstrates the EE convergence performance under different numbers of the multi-head attention $N_H$ and dropout rates $D_r$ applied in the deep neural networks in the proposed SOHRL framework. We can infer that the case with $N_H=4$ achieves the highest EE which is close to $0.7$ bits/Hz/J with $D_r=0.5$ and slightly lower with $D_r=0.3$. This indicates that a moderate number of attention heads can effectively capture diverse feature representations while avoiding excessive redundancy of hidden information. By contrast, employing $N_H=2$ heads in attention leads to a lower EE around $0.4$-$0.45$ bits/Hz/J. This is because the reduced number of attention modules limits the model ability to extract sufficient feature correlation from the input states. On the other hand, $N_H=8$ heads instead results in the lowest EE, as it can be attributed to potential overfitting issue, i.e., excessive heads might weaken meaningful attention weights and introduce instability during learning process. Moreover, it shows that dropout $D_r=0.5$ has lower EE compared to that of $D_r=0.3$ in all cases. This suggests that while dropout prevents overfitting, excessive dropout of neural weights disrupts the hidden information extraction and slow down the convergence. The EE difference between droprate settings is most pronounced when $N_H=4$, striking the compelling trade-off between parameter regularization and feature expressiveness for the achieving maximum EE performance.

\begin{figure}[!t]
\centering
	\subfigure[]{
        \includegraphics[width=1.6in]{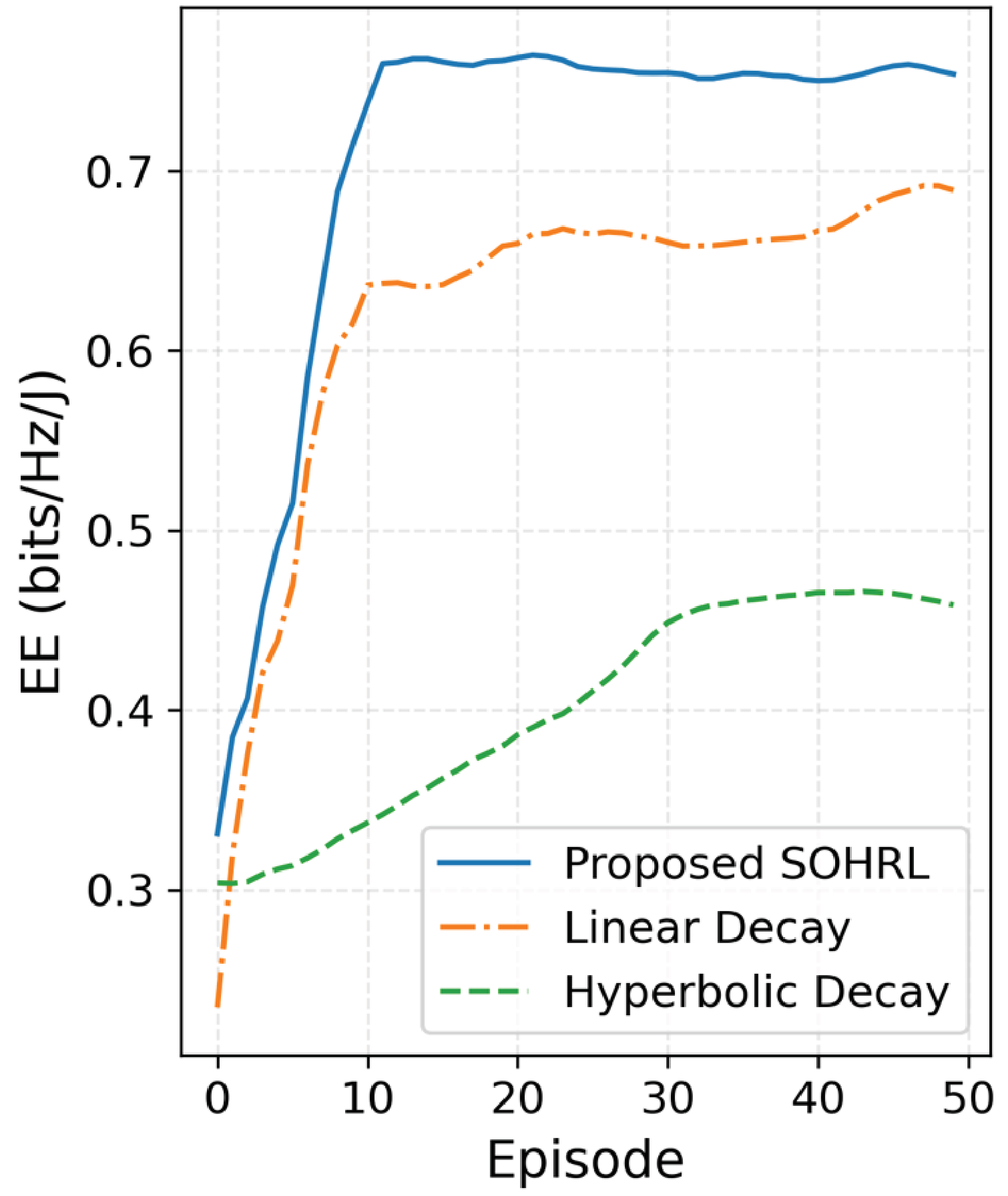} \label{fig_eps1}}
        \subfigure[]{
        \includegraphics[width=1.6in]{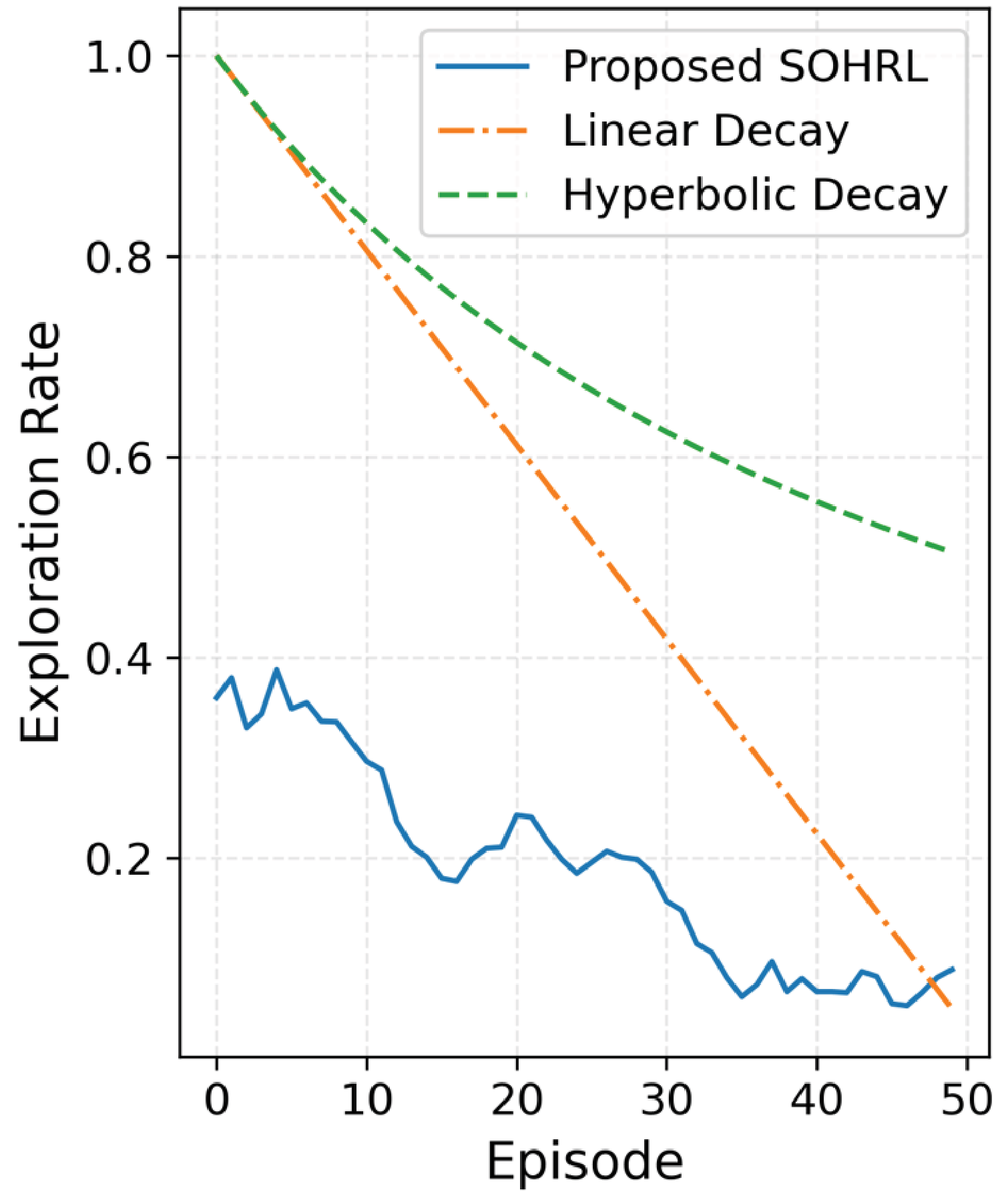} \label{fig_eps2}}       
    \caption{Convergence of different exploration methods in MADQN in terms of (a) EE and its corresponding (b) exploration rate.}
    \label{fig_eps}
\end{figure}

Fig.~\ref{fig_eps} depicts the effect of different exploration methods in MADQN, including the proposed SOHRL with meta-PPO, linear decay and hyperbolic decay of exploration rates. Note that linear decay indicates that $\varsigma$ decays uniformly and linearly from $1$ to $0.05$, whereas hyperbolic decay adopts the decaying speed of $\varsigma \leftarrow \varsigma/t$. As observed from Fig. \ref{fig_eps1}, the proposed SOHRL scheme achieves the highest EE performance owing to its dynamic adjustment of the exploration rate. The linear decay method exhibits moderate EE performance as it relies on a wide range of the exploration rate while failing to account for instantaneous performance factors. In contrast, the hyperbolic decay approach has the lowest EE performance, since it requires more training epochs to acquire high-quality solutions under limited exploration-rate variations. Moreover, Fig. \ref{fig_eps2} indicates that both SOHRL and the linear decay method employ relatively large exploration rates during the initial learning to facilitate solution search, which is followed by lower exploration rates to enhance learning stability and convergence.

\begin{figure}[!t]
\centering
	\subfigure[]{
        \includegraphics[width=3in]{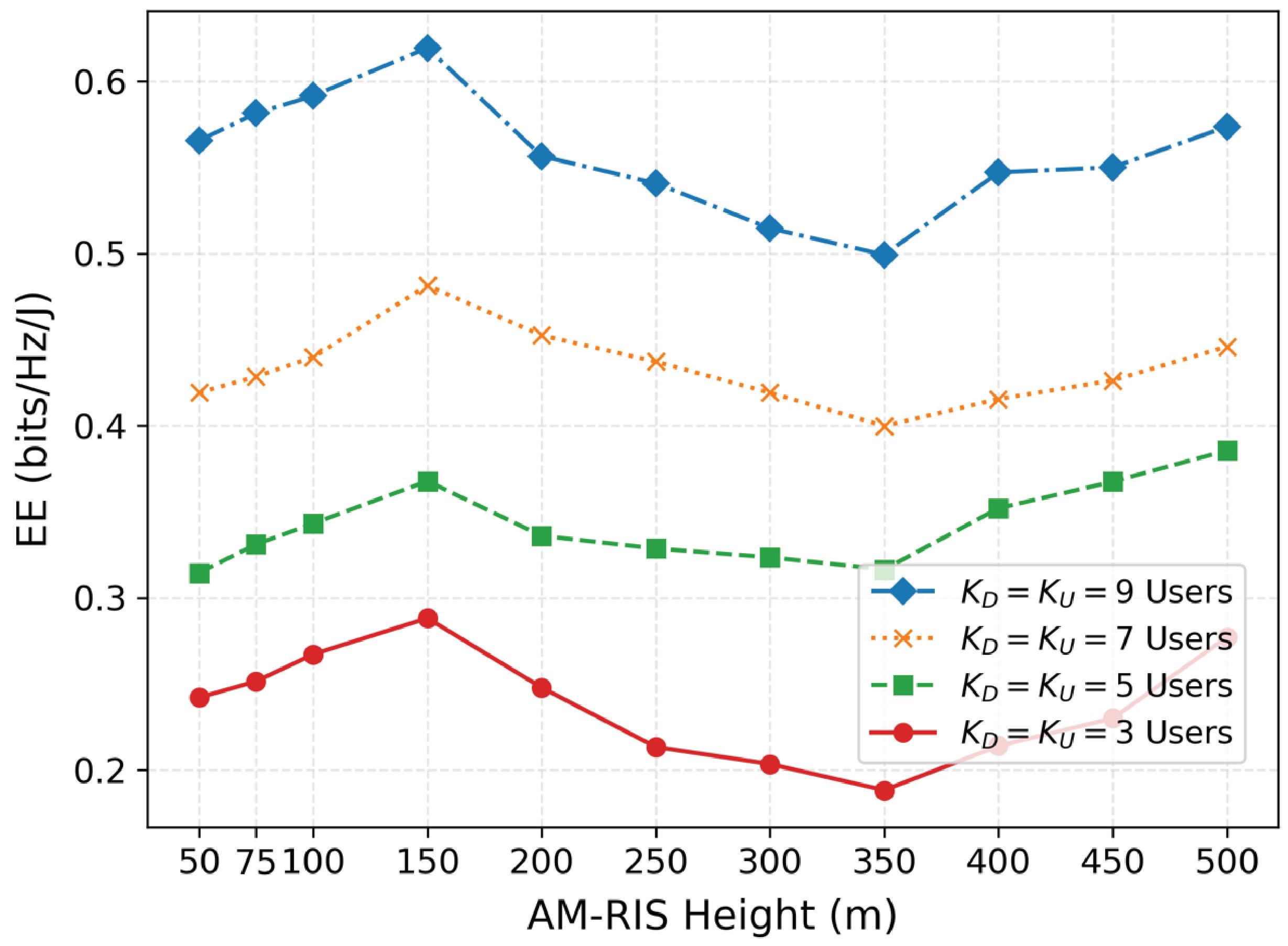} \label{fig4a}}
        \subfigure[]{
        \includegraphics[width=3in]{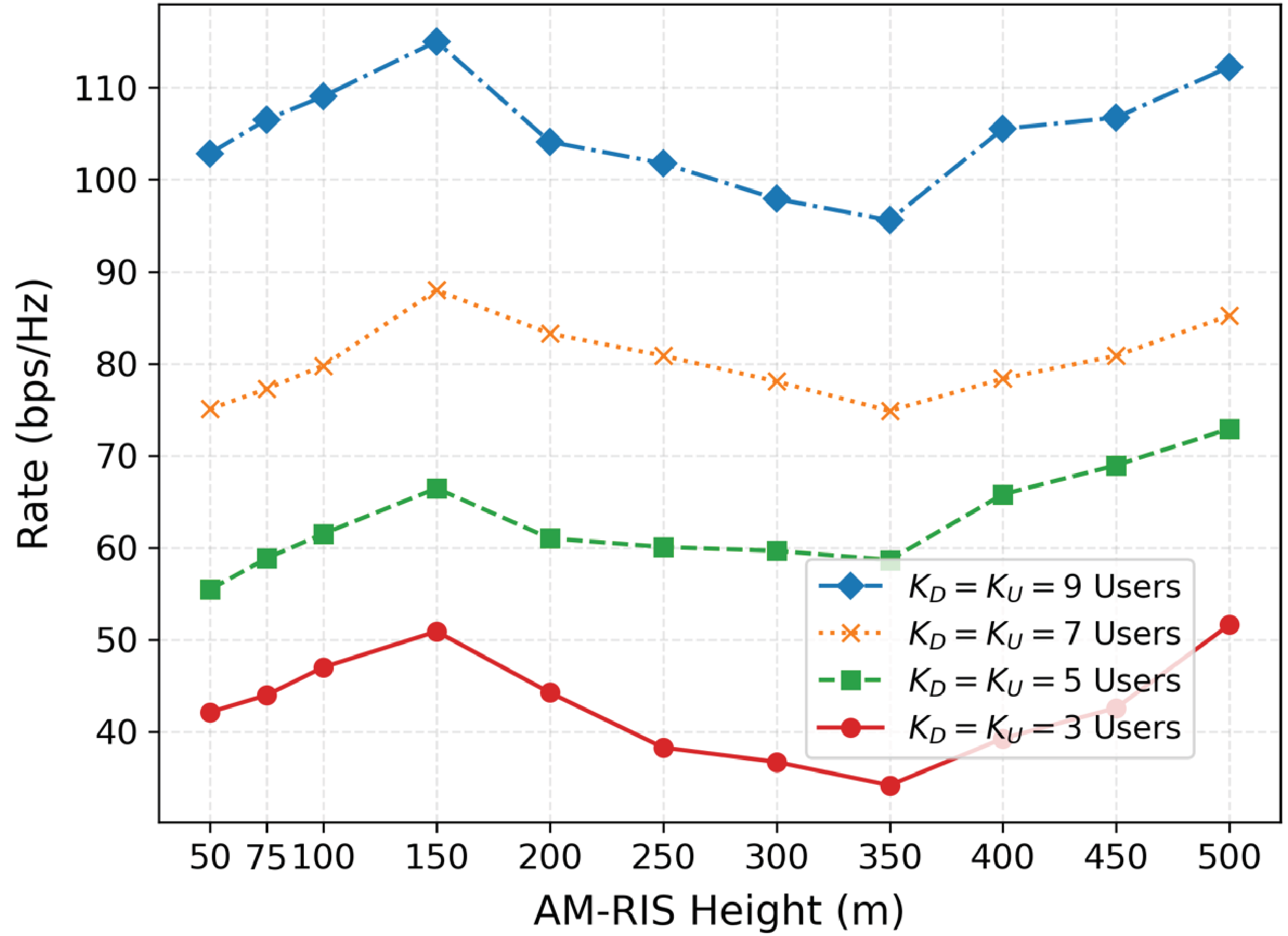} \label{fig4b}}
	\subfigure[]{
        \includegraphics[width=3in]{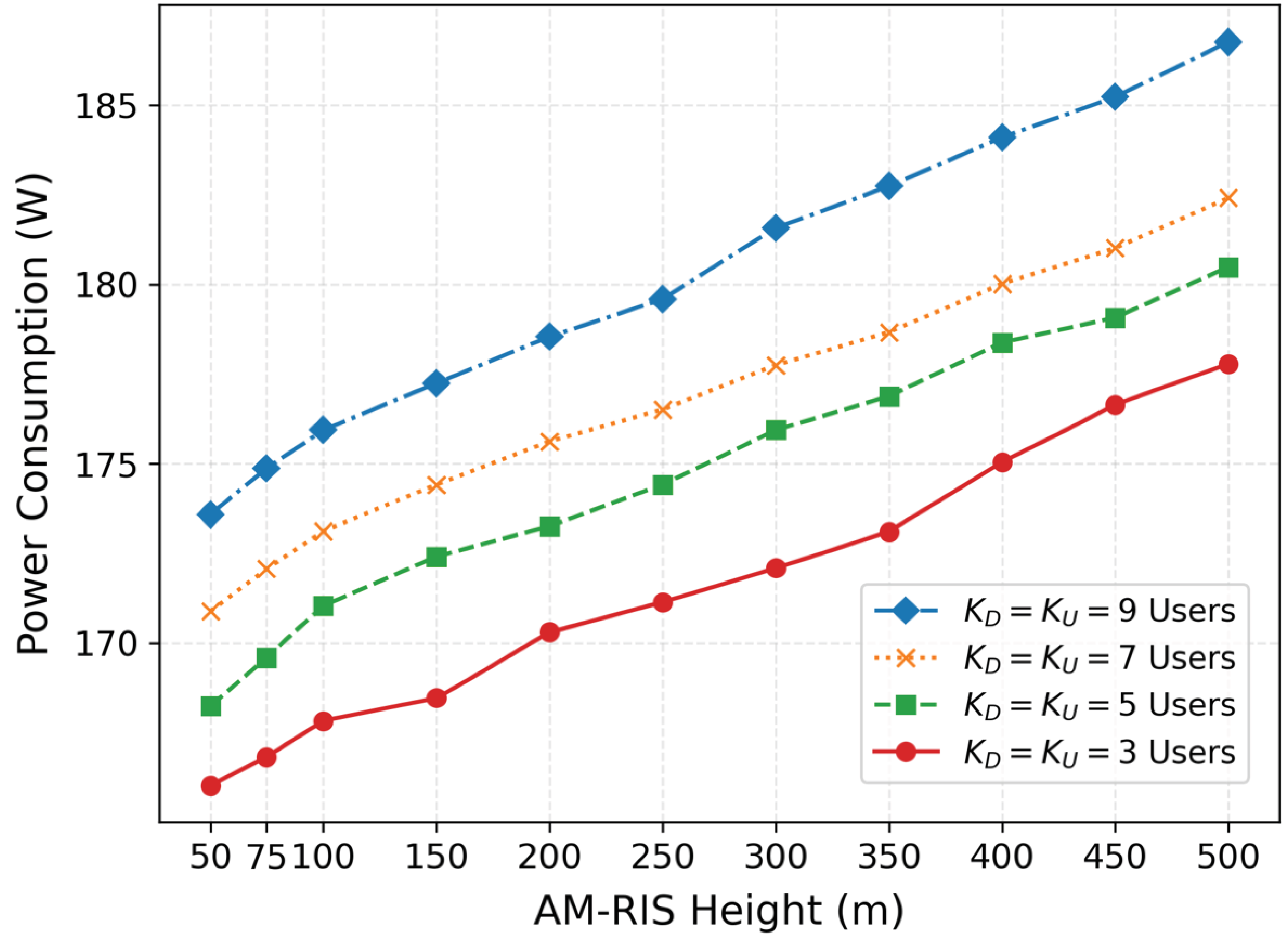} \label{fig4c}}        
    \caption{EE performance with different deployment heights of AM-RISs and various numbers of users w.r.t. (a) EE, (b) rate, and (c) power consumption.}
    \label{fig4}
\end{figure}

Fig.~\ref{fig4} illustrates the effect of AM-RIS deployment height on the EE performance in terms of EE, rate, and power consumption respectively in Figs. \ref{fig4a}, \ref{fig4b}, and \ref{fig4c} under different numbers of uplink and downlink users with $K_{\rm D}=K_{\rm U}$. We can observe that both EE and rate exhibit a non-monotonic trend w.r.t. height. The EE curves initially increase as the height grows and reach the maximum EE at the height of around $150$ m. At the heights of $[150, 350]$ m, EE gradually decreases due to the dominant path loss over the benefits of improved coverage. In contrast, the power consumption monotonically increases with the height for all user scenarios, rising from around $168$ W at a height of $50$ m to more than $185$ W at that of $500$ m, with more users leading to larger power consumption. Moreover, more users significantly enhance both EE and rate thanks to the multiplexing gains. Intriguingly, at the heights of $[350, 500]$ m, EE slightly improves again. Such counter-intuitive trend can be explained by the interplay between the rate and power consumption of AM-RISs. At higher deployed heights, the AM-RIS can provide broader coverage with more favorable incidence and reflection angles, which reduces interference. While the path loss indeed grows with height, the system benefits from the improved spatial diversity of AM-RIS and the reduced blockage effects. These results indicate that a moderate deployment height at around $150$ m strikes the compelling trade-off between the coverage and attenuation, maximizing EE and rate.

\begin{figure}[t]
    \centering
    \includegraphics[width=3in]{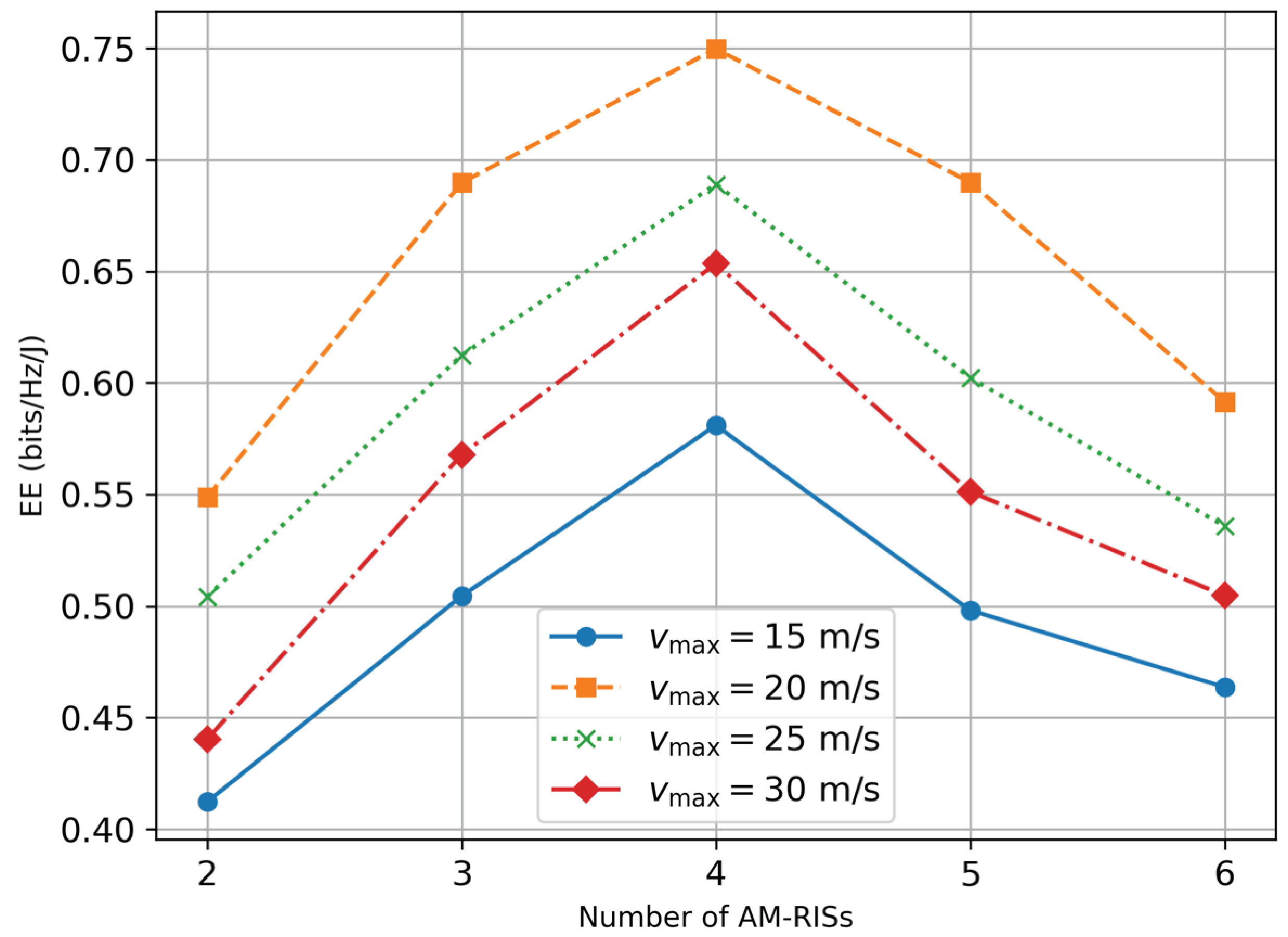}
    \caption{EE performance under different numbers of AM-RISs and the maximum velocity values of AAVs.}
    \label{fig5}
\end{figure}

Fig. \ref{fig5} presents the EE performance with different numbers of AM-RISs under different maximum velocity of $v_{\max}$. We can observe that EE first increases with increment of the number of AM-RISs and then decreases after $4$ AM-RISs. This concave-shaped curve indicates that adding more AM-RISs initially enhances spatial diversity and improves reflective channel conditions, whilst excessive AM-RIS deployment leads to more power consumption, thereby diminishing the overall EE. In terms of mobility, moderate velocity of $v_{\max}=20$ m/s achieves the highest EE, reaching around $0.75$~bits/Hz/J with $4$ AM-RISs. When the AM-RIS is slow $v_{\max}=15$ m/s, the system lacks sufficient adaptability in learning dynamic environments, leading to reduced EE. Conversely, at higher velocity values of $v_{\max}\in\{25,30\}$ m/s, EE decreases owning to more frequent channel variations, limiting the stability of the learning process.

\begin{figure}[t]
    \centering
    \includegraphics[width=3in]{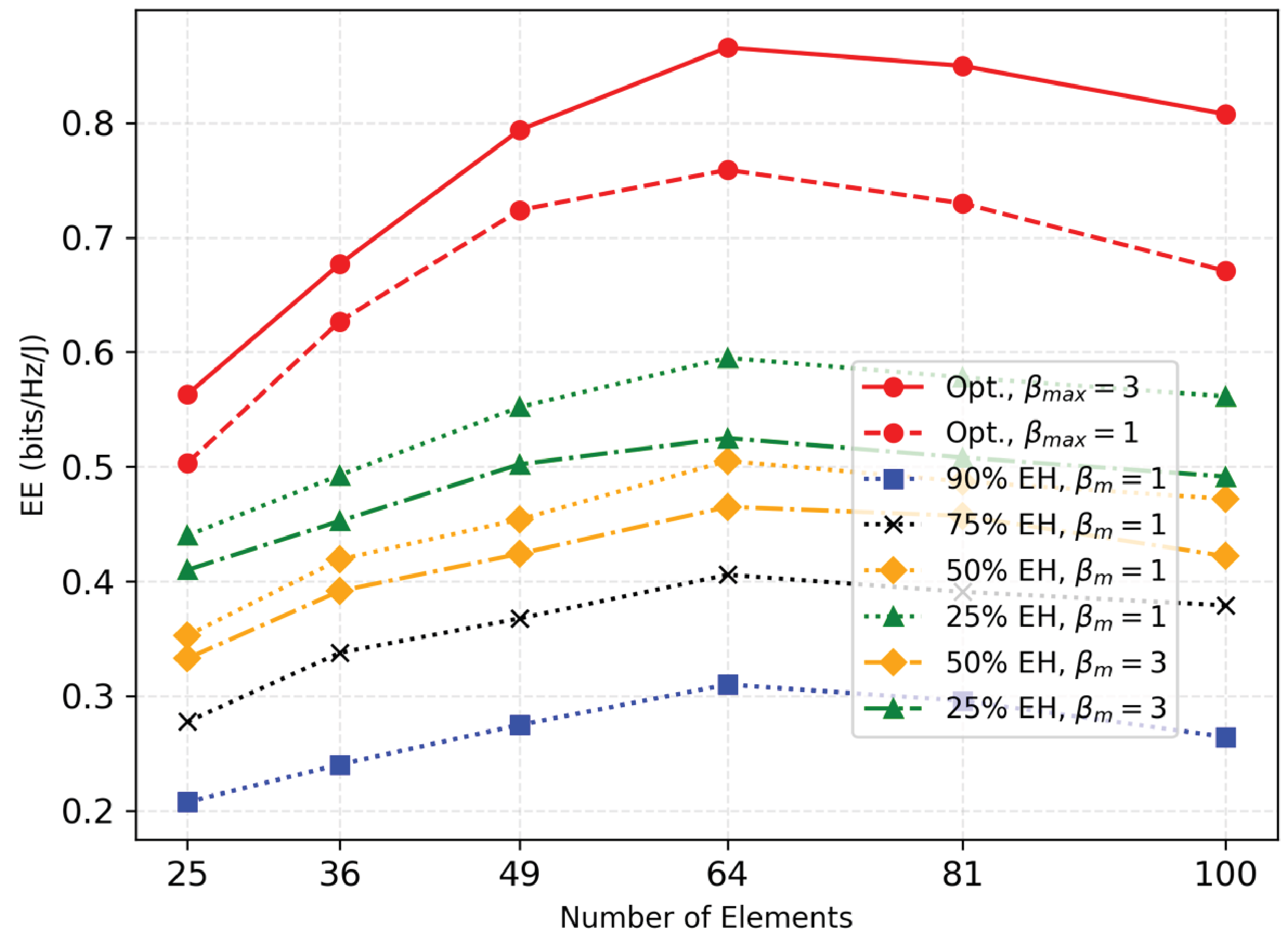}
    \caption{Comparison the Energy Efficency performance with different numbers of MF-RIS elements.}
    \label{fig6}
\end{figure}

Fig.~\ref{fig6} illustrates the EE performance under various numbers of AM-RIS elements under different configurations of EH ratio and maximum amplitude factors $\beta_{\max}$. Here, the EH ratio denotes the portion of randomly selected AM-RIS elements dedicated solely to EH capability. Note that $\beta_{\max}=1$ corresponds to the passive AM-RIS operation and $\beta_{\max}>1$ indicates active AM-RIS with amplification capability. We can observe that the optimal case with full optimization accomplishes the highest EE performance, particularly when $\beta_{\max}=3$, reaching nearly $0.84$ bits/Hz/J with $64$ AM-RIS elements. This demonstrates the significant performance gain benefited from the active element with amplification function, as larger amplitude flexibility enhances the effective reflective signal strength. Even under the passive configuration with $\beta_{\max}=1$, the optimal design still provides higher EE than the partial EH cases. As the EH ratio increases, EE degrades because fewer elements are available for reflection and beamforming. For instance, the case with $90\%$ of elements with EH performs the worst EE of around $0.3$~bits/Hz/J, as most elements cannot contribute to the reflection function. Moderate EH ratios of $25\%$ and $50\%$ strike a compelling balance between enabling significant reflection gain and harvesting energy, achieving EE values of around $0.5$-$0.6$ bits/Hz/J. Additionally, the EE performs a concave shape w.r.t. the number of AM-RIS elements. Furthermore, the EE performance increases with the number of elements and peaks at $64$ elements, but slightly decreases beyond it. This is due to reason that the diminishing marginal beamforming gain and increasing power consumption of AM-RISs as more elements are deployed.

\begin{figure}[t]
    \centering
    \includegraphics[width=3in]{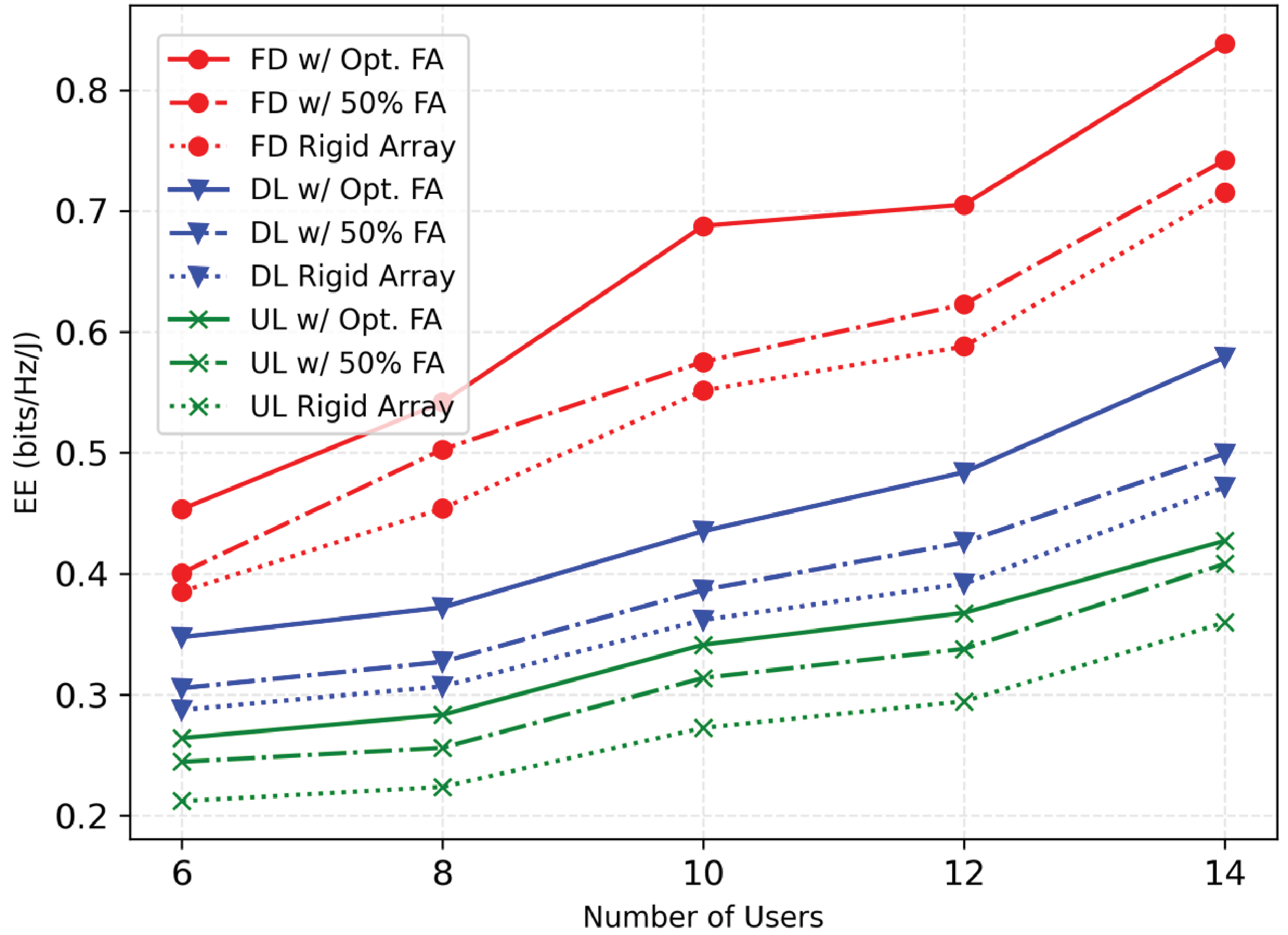}
    \caption{EE performance of different configurations of optimized FA, partial FA ($50\%$) and conventional rigid array in FD and HD-based DL/UL under various numbers of users.}
    \label{fig9}
\end{figure}

Fig. \ref{fig9} depicts the EE performance versus the number of users under different transmissions and FA configurations. Note that DL and UL correspond to scenarios with only DL and UL users in HD mode, respectively. The antenna architectures include: (1) "Opt. FA" for fully optimized fluid antenna configurations; (2) "$50\%$ FA" with only half of the antennas are fluid and the others remained fixed in a conventional rigid array; and (3) "Rigid Array" refers to a fully fixed antenna structure with equal inter-antenna spacing. It can be observed that FD with optimized FA configuration achieves the highest EE across all numbers of users, reaching up to $0.82$ bits/Hz/J when $14$ users. This is benefited by the simultaneous exploitation of FD transmissions along with the spatial flexibility of FA, enabling interference mitigation and adaptive beamforming. In contrast, FD with $50\%$ FA or with rigid arrays shows a slight degradation in EE, highlighting the importance of full FA reconfigurability. For the HD scenarios, the DL case outperforms the UL one in EE performance, since the DL transmissions benefit from more available power budget at BS. Nevertheless, both DL and UL achieve significantly lower EE than FD due to the lack of simultaneous transmissions. Similar to FD, employing optimized FA in DL and UL yields substantial EE gains compared to cases of $50\%$ FA and rigid arrays. While, the rigid array structure providing limited flexibility remains the least energy efficient across all cases.

\begin{figure}[t]
    \centering
    \includegraphics[width=3in]{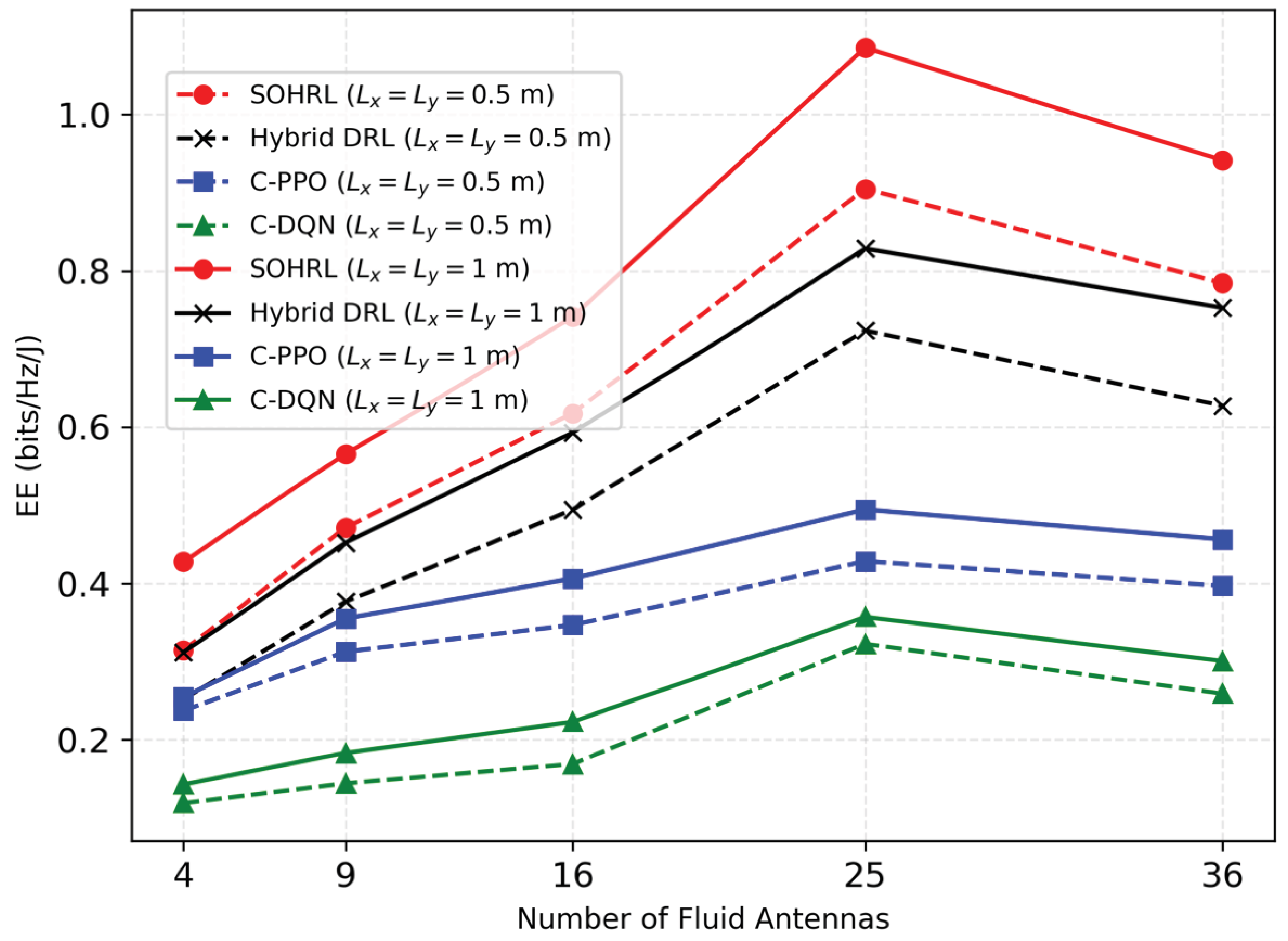}
    \caption{Comparison of four schemes under different fluid-antenna array sizes}
    \label{fig10}
\end{figure}

Fig.~\ref{fig10} compares the EE performance versus the number of FAs under different DRL schemes and antenna array sizes. It can be observed that SOHRL outperforms all benchmarks across both antenna array size thanks to hybrid PPO-DQN framework with attention-driven state representations and hyperparameter self-optimization. With $25$ FAs\textsuperscript{\ref{note3}}\footnotetext[3]{For sub-6 GHz deployments, a movable range of 0.1 m corresponds to few-wavelength aperture in existing fluid/movable antenna studies. Such a compact structure is suitable for the user side. In contrast, the larger FA structure with the size of $L_{x/y} \in \{0.5, 1 \} $ m is intended for the BS side, which is physically feasible and aligns with the form factor of BS antenna arrays. The large-aperture spans multiple wavelengths, capable of providing larger spatial degrees of freedom. \label{note3}} and $L_x=L_y=1$ m, SOHRL achieves an optimal EE around $0.11$ bits/Hz/J, whereas hybrid DRL, C-PPO, and C-DQN peak at about $0.81$, $0.5$, and $0.36$ bits/Hz/J, respectively. This demonstrates that the joint design of attention and hyperparameter optimization is essential for capturing unequal importance of input states. In contrast, hybrid DRL without these mechanisms suffers from reduced efficiency, whereas the centralized single-neural network baselines fail to learn strategy effectively due to the high-dimensional action space. With different numbers of FAs, EE improves up to around $25$ antennas thanks to improved spatial diversity gains, which is followed by the declined EE due to increased power consumption of operating more antennas. Additionally, larger array size of $L_x=L_y=1$ m is capable of containing more FAs, providing higher EE than the smaller array size of $L_x=L_y=0.5$ m, as the former enables higher spatial degree of freedom of adjustment of antenna positions for combating small-scale channel fading.

\begin{figure}[t]
    \centering
    \includegraphics[width=3in]{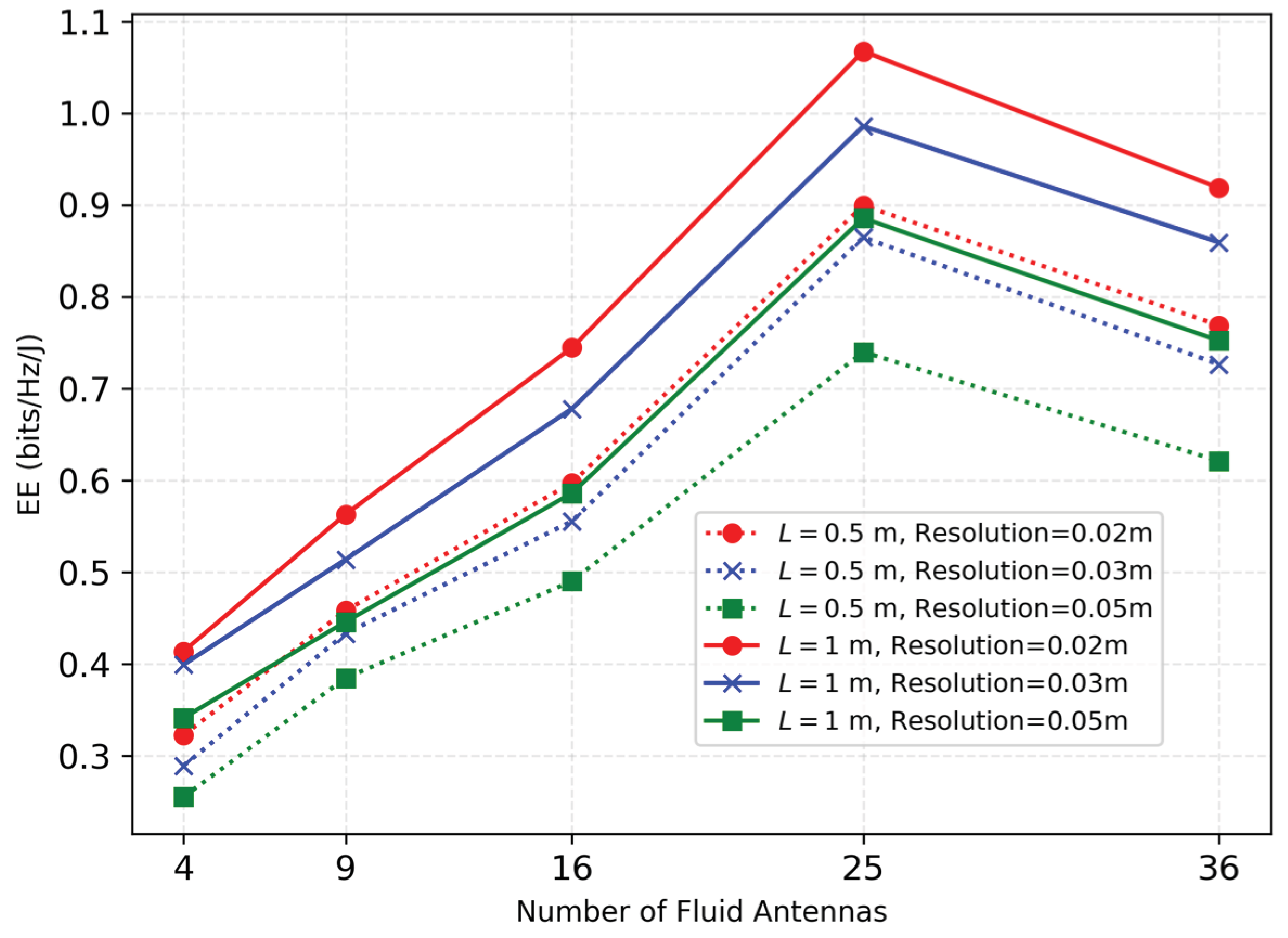}
    \caption{EE performance of different FA array sizes of $L\in\{0.5,1\}$ m and resolutions of $\{0.02, 0.03, 0.05\}$ m.}
    \label{fig_qp}
\end{figure}

Fig.~\ref{fig_qp} compares the EE performance under different FA array lengths $L_x=L_y=L\in\{0.5,1\}$ m and spatial resolutions of $\{0.02, 0.03, 0.05\}$ m. It is observed that finer spatial resolutions of $0.02$ m outperform coarser ones, as they provide a denser set of candidate FA positions for exploiting favorable channel realizations. Moreover, the FA configuration with a resolution of $0.02$ m under $L=0.5$ m achieves higher EE than that with a coarser resolution of $0.05$ m and a larger array length of $L=1$ m. This indicates that increasing the spatial resolution is more effective than merely enlarging the array size in leveraging favorable channel conditions. Consequently, achieving the optimal EE performance requires a compelling tradeoff among the FA array size, spatial resolution of FA ports, and the number of antennas.

\begin{figure}[t]
    \centering
    \includegraphics[width=3in]{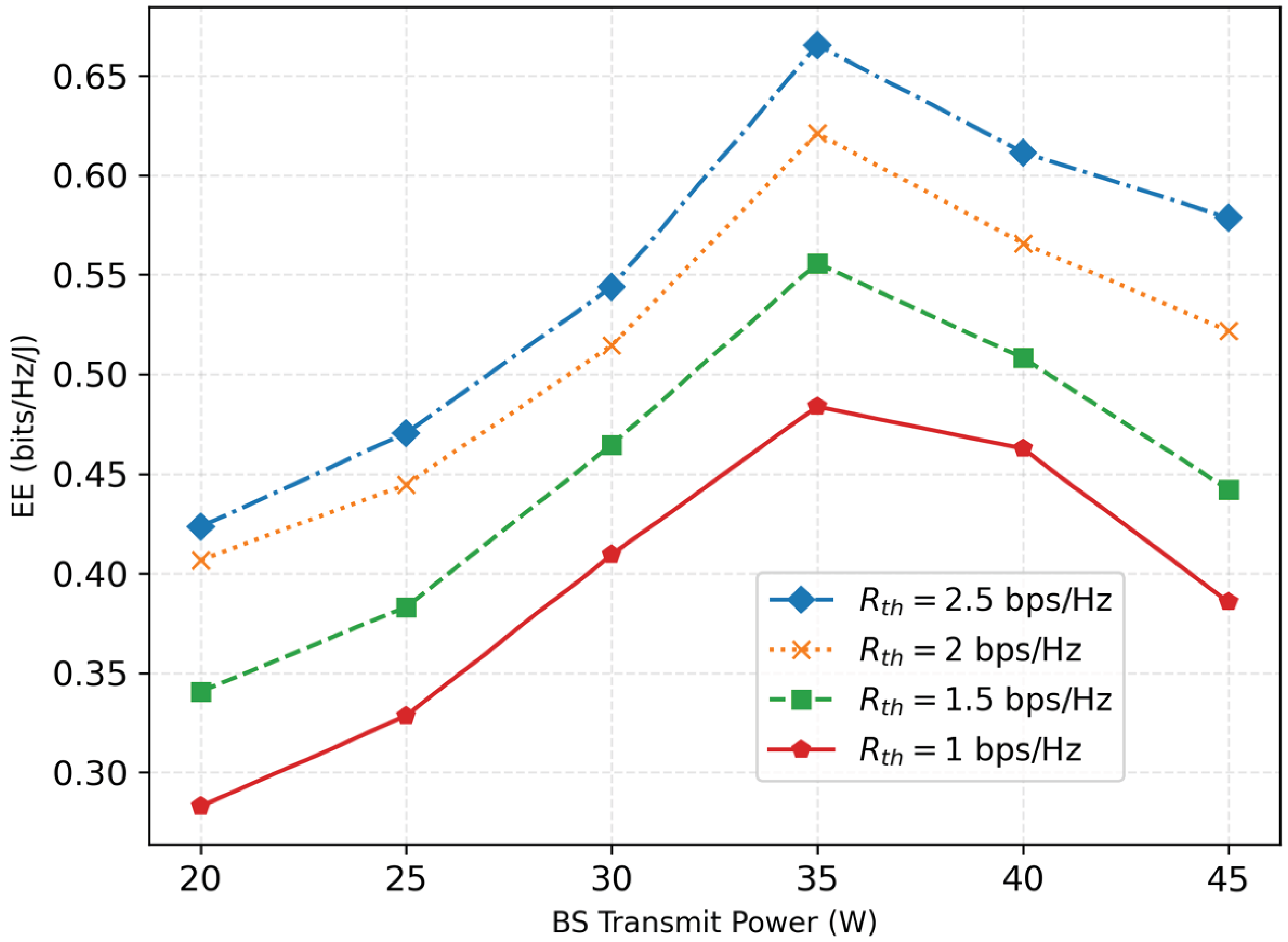}
    \caption{Comparison of four schemes under different fluid-antenna array sizes}
    \label{fig11}
\end{figure}

Fig.~\ref{fig11} illustrates the EE performance w.r.t. the maximum BS transmit power under different minimum user rate requirements $R_{\mathrm{th}}$. It can be observed that the EE first increases with transmit power and then decreases after a peaked EE around $P_{\mathrm{BS}}=35$ W. Such trend arises from the trade-off between throughput improvement and power consumption. Increasing $P_{\mathrm{BS}}$ enhances the achievable rate due to more available power resources to be allocated, but excessive power consumption potentially leads to a reduced EE. Comparing different rate requirements, higher $R_{\mathrm{th}}$ values provide better EE performance, i.e., when $R_{\mathrm{th}}=2.5$ bps/Hz, the system achieves a highest EE around $0.67$ bits/Hz/J, while the optimal EE is around $0.48$ bits/Hz/J when $R_{\mathrm{th}}=1$ bps/Hz. This is because stricter rate constraints push the learning strategy to more aggressively optimize resource allocation, thereby improving EE.

\begin{table}[!t]
\centering
\scriptsize
\setstretch{1.25}
\caption{Computational Complexity}
\begin{tabular}{|l|l|}
\hline
\textbf{Algorithm} & \textbf{Computational Complexity} \\
\hline
C-DQN & 
$\mathcal{O}\left( N_B N_L N_W^2  + N_B  N_W \cdot |\mathcal{A}^{p}_{\mathcal{Q}}| \right)$ \\
\hline
C-PPO & 
$\mathcal{O}\left( N_B N_L N_W^2 + N_B N_W \cdot |\mathcal{A}^{p}_{\mathcal{C}}| \right)$  \\
\hline
MAPPO & 
$\mathcal{O}\left( N_B N_L N_W^2 + N_B N_W \cdot |\mathcal{A}^{p}_{i, \mathcal{C}}| \right)$ \\
\hline
Hybrid DRL & 
$\mathcal{O} \Big( 2 N_B N_L N_W^2 + N_B N_W \cdot (|\mathcal{A}^{\text{c}}| + |\mathcal{A}^{\text{d}}|) \Big)$ \\
\hline
\tabincell{l}{Proposed SOHRL \\ w/o Hyperparameter \\ Optimization} & \tabincell{l}{$\mathcal{O} \Big( 2 N_B N_L N_W^2 + N_B N_W \cdot (|\mathcal{A}^{\text{c}}_i| + |\mathcal{A}^{\text{d}}_i|)$ \\ \quad $+N_B \cdot (4 |\mathcal{S}_i| N_W^2 N_H) \Big)$} \\
\hline
\tabincell{l}{Proposed SOHRL \\ w/o Attention} & $\mathcal{O} \Big( 3 N_B N_L N_W^2 + N_B N_W \cdot (|\mathcal{A}^{\text{c}}_i| + |\mathcal{A}^{\text{d}}_i| + |\mathcal{G}|) \Big)$ \\
\hline
Proposed SOHRL & \tabincell{l}{$\mathcal{O} \Big( 3 N_B N_L N_W^2 + N_B N_W \cdot (|\mathcal{A}^{\text{c}}_i| + |\mathcal{A}^{\text{d}}_i| + |\mathcal{G}|)$ \\ \quad $+N_B \cdot (4 |\mathcal{S}_i| N_W^2 N_H) \Big)$} \\
\hline
\end{tabular}
\label{complexity}
\end{table}

Table~\ref{complexity} summarizes the computational complexity of the proposed SOHRL scheme and its simplified variants as well as the benchmark methods. Let $N_B$ denote the training mini-batch size, $N_L$ the number of neural network layers, and $N_W$ the number of hidden neurons per layer. The notations $\mathcal{A}^{p}_{\mathcal{Q}}$, $\mathcal{A}^{p}_{\mathcal{C}}$, and $\mathcal{A}^{p}_{i,\mathcal{C}}$ represent the quantized joint action set, the continuous joint action set, and the continuous action set associated with the $i$-th agent, respectively. Simulations are conducted with an Intel Core i7-13700 central processing unit (CPU) and NVIDIA GeForce RTX 3060Ti graphics processing unit (GPU). Each agent driven by C-DQN, C-PPO, MAPPO, hybrid DRL, SOHRL without hyperparameter optimization, SOHRL without attention, and the proposed SOHRL scheme requires storage memories of $\{2.6, 2.8, 2.9, 5.4, 4.6, 5.1, 6.2\}$ MB according to the parameter count, and incurs backward-propagation training time of $\{27, 32, 40, 46, 48, 52, 68\}$ ms per training step. The inference overhead is negligible compared with the training phase, as it only involves forward propagation. Although C-DQN achieves the lowest overhead, its limited flexibility and large quantization errors in handling hybrid continuous-discrete actions restricts its learning capability in complex and high-dimensional systems. The results reveal that the proposed SOHRL scheme provides practical gains in system EE improvement with moderate computational overhead, making it well suited for future AM-RIS-assisted and FA-aided FD networks.

\section{Conclusions} \label{sec_con}

We have conceived a novel architecture of multi-AM-RISs for FA-aided FD network, where each AM-RIS is implemented on AAV via MF-RIS. The formulated problem aims at maximization of system EE, guaranteeing constraints of available operating power, AM-RIS configurations and flight requirement, as well as the deployable FA and AAV region. To address this non-solvable problem, we have proposed a SOHRL learning framework by employing a hybrid DRL architecture, where MADQN deals with the discrete actions and MAPPO tackles continuous cases. Moreover, attention-driven states provide unequal importance weight of inputs, whilst the meta-PPO supports the self-optimization and auto-tuning of hyperparameters in DRL. The results have demonstrated the effectiveness of the proposed architecture empowered by SOHRL. The results reveal that SOHRL outperforms existing benchmarks of the case without attention mechanism and conventional hybrid/multi-agent/standalone DRL. Moreover, the proposed AM-RIS in FD architecture achieves the highest EE performance compared to half-duplex, conventional rigid antenna arrays, partial EH, and conventional RIS without amplification, highlighting its potential as a compelling solution for EE-aware wireless networks.

\bibliographystyle{IEEEtran}
\bibliography{IEEEabrv}

\end{document}